\documentclass[twocolumn]{aastex63}

\received{2022 August 24}
\accepted{2023 February 6}
\submitjournal{PASP}

\shorttitle{Comparison of TDE models with TiDE}
\shortauthors{Kov\'acs-Stermeczky \& Vink\'o}

\begin{document}

\title{Comparison of different Tidal Disruption Event light curve models with TiDE, a new modular open source code}

\correspondingauthor{Zs. Kov\'acs-Stermeczky}
\email{stermeczky.zsofia@csfk.org}

\author[0000-0002-5655-0154]{Zs\'ofia V. Kov\'acs-Stermeczky}
\affiliation{ELTE E\"otv\"os Lor\'and University, Institute of Geography and Earth Sciences, Department of Astronomy, Budapest, Hungary}

\affiliation{ Konkoly Observatory,  CSFK, Konkoly-Thege M. \'ut 15-17, 
Budapest, 1121, Hungary}

\author[0000-0001-8764-7832]{J\'ozsef Vink\'o}
\affiliation{ Konkoly Observatory,  CSFK, Konkoly-Thege M. \'ut 15-17, 
Budapest, 1121, Hungary}
\affiliation{ELTE E\"otv\"os Lor\'and University, Institute of Physics, P\'azm\'any P\'eter s\'et\'any 1/A, Budapest, 1117 Hungary}
\affiliation{Department of Astronomy, University of Texas at Ausin, 2515 Speedway Stop C1400, Austin, TX, 78712-1205, USA}
\affiliation{Department of Experimental Physics, University of Szeged, D\'om t\'er 9, Szeged, 6720, Hungary}

\begin{abstract}
A tidal disruption event (TDE) occurs when a supermassive black hole disrupts a nearby passing star by tidal forces. The subsequent fallback accretion of the stellar debris results in a luminous transient outburst. Modeling the light curve of such an event may reveal important information, for example the mass of the central black hole. This paper presents the TiDE software based on semi-analytic modeling of TDEs. This object-oriented code contains different models for the accretion rate and the fallback timescale $t_{\rm min}$. We compare the resulting accretion rates to each other and with hydrodynamically simulated ones and find convincing agreement for full disruptions. We present a set of parameters estimated with TiDE for the well-observed TDE candidate AT2019qiz, and compare our results with those given by the MOSFiT code. Most of the parameters are in reasonable agreement, except for the mass and the radiative efficiency of the black hole, both of which depend heavily on the adopted fallback accretion rate.
\end{abstract}


\section{Introduction} \label{sec:intro}

\hspace{0.5 cm}

In a Tidal Disruption Event (TDE) a supermassive black hole (BH) disrupts a nearby passing star by tidal forces. This phenomenon was predicted theoretically in the 1970s \citep{1975Natur.254..295H}, and many studies have appeared in the literature since then. Recent reviews of this popular topic can be found in \citet{2020SSRv..216..124V}, or in \citet{2021ARA&A..59...21G}.

During a TDE a star approaches the black hole closer than a critical distance (the tidal radius), and getting disrupted by the gradient of the gravitational force from the BH. In a first approximation, about half of the stellar debris remains bound to the BH, while the other half leaves the system \citep{1988Natur.333..523R}. The bound material starts falling back to the pericenter, and quickly forms an accretion disk around the BH. 

The light curve of such a system consists of emission from at least two components: an accretion disk and a super-Eddington wind \citep[e.g.][]{Strubbe09}. Early implementations of TDE light curve models \citep[e.g.][]{Lodato11} were built assuming that these two components are the dominant ones. Since then several other mechanisms have been proposed, including the reprocessing of the high-energy radiation from the disk to optical photons by the wind material \citep[e.g.][]{Guillochon14}, the effect of shocks near the apocenter of the infalling stream \citep{2015ApJ...804...85S, 2015ApJ...806..164P} or shocks due to stream-disk interactions \citep{2020MNRAS.495.1374B, 2022arXiv220610641S} that may also power the light curve around the initial peak. 

The rich physics involved in TDEs have been investigated from many different aspects.  For example,  \citet{2016MNRAS.463.2242X} divides TDE-related studies into four groups. Papers in the first group examine the dynamical properties, i.e. how a star can approach the BH close enough, like e.g. in \citet{2016MNRAS.455..859S}.
The reason why stars go into disruptive orbits is still not clear, but there are several theories from the standard two body relaxation \citep{1976MNRAS.176..633F,1999MNRAS.309..447M,2016MNRAS.455..859S} to the presence of SMBH binaries \citep{2005MNRAS.358.1361I,2009ApJ...697L.149C,2011ApJ...729...13C,2011MNRAS.412...75S,2015MNRAS.451.1341L}. From these scenarios it seems possible that the orbit of the star can also be highly eccentric instead of parabolic, even though the fraction of TDEs from such eccentric systems may be low compared to the number of close encounters from parabolic orbits. If the star can survive the approach, it could exhibit periodic outbursts \citep{2013ApJ...777..133M,2022ApJ...927L..25N}.
In \citet{2022arXiv220613494L} the authors showed that stars on highly eccentric orbits can produce multiple, short-lived TDEs compared to parabolic encounters, which may explain peculiar, repeating TDEs like ASASSN-14ko.

The second group of papers focuses on the TDE phenomenon itself via semi-analytic calculations or hydrodynamical simulations \citep{lodato09,2013ApJ...767...25G, 2016MNRAS.461..948M, Golightly_2, Golightly_1, nixon21}. Studies in the third group investigate the physics and evolution of the accretion disk formed by the debris of the disrupted star, see e.g. \citet{Hayasaki13}, \citet{Dai15}, \citet{2015ApJ...806..164P}, \citet{Hayasaki16}, \citet{2016MNRAS.455.2253B},  \citet{2020MNRAS.495.1374B} or \citet{2022arXiv220610641S}. The fourth group explores the observable properties of TDEs. Reviews of the optical observations can be found e.g. in \citet{2020SSRv..216..124V} and \citet{2021ARA&A..59...21G}, while for X-ray bright TDEs see e.g. \citet{Komossa15} and \citet{Saxton20}.  

During the last few years the number of observations of TDE candidates has grown significantly. These events were studied both via photometry and spectroscopy. Many optical observations were produced by large surveys, like the Sloan Digitized Sky Survey (SDSS) \citep[e.g.][]{2008ApJ...678L..13K, 2011ApJ...741...73V, 2011ApJ...740...85W, 2012ApJ...749..115W, 2013ApJ...774...46Y} or the Zwicky Transient Facility (ZTF) \citep{2019ApJ...872..198V, 2021ApJ...908....4V}. For fitting the light curve of a TDE many authors used the popular open source MOSFiT code \citep[e.g.][]{2019ApJ...872..151M, 2020MNRAS.499..482N, 2022arXiv220613494L}

In this paper we concentrate on semi-analytical modeling of the UV/optical contribution to the observable radiation from the disruption of a stellar-mass encounter by a supermassive BH. 

For the basic dynamical and radiative transfer processes we mostly follow the studies by \citet{Strubbe09}, \citet{lodato09} (L09 hereafter) \citet{Lodato11} and \citet{Guillochon} (GR13). We also discuss the model described by \citet{2019ApJ...872..151M}, which is implemented in the widely used MOSFiT code \citep{2018ApJS..236....6G}. 

The main purpose of our paper is to introduce TiDE, an open-source, publicly available code that can be applied to fit the UV/optical photometry of TDEs effectively. Even though the TDE module implemented in MOSFiT became a de facto standard for this purpose, our model is based on slightly different physical assumptions, thus, it likely produces somewhat different results for the same data. Thus, it may be used to explore the possible parameter space in more detail, and may reduce the systematic biases in the posterior probabilty distributions introduced by the model assumptions.  

This paper is organized as follows. The main assumptions for calculating a TDE light curve are summarized in Section \ref{sec:tde_model}. The available models for the accretion rate and their comparison with hydrodynamically simulated ones (GR13) are presented in Section \ref{sec:different_models}. In Section \ref{sec:light_curves_comparsion} we examine the effect of different parameters on the light curves and the decline rates. We also demonstrate the limits of a TDE event involving a white dwarf (WD) star and visualize the differences between the disruption of a WD and main sequence (MS) star. Finally, in Section \ref{sec:at2019qiz} we test the predictions of our model by fitting the UV/optical observations of AT2019qiz, and compare the resulting physical parameters with those given by the MOSFiT code.

\section{The TDE Model}
\label{sec:tde_model}
In this section we summarize the equations that were used in computing the TDE model light curves. Most of these are based on papers by \citet{Lodato11} and \citet{Strubbe09} paper. In this section we focus on those equations that have exact forms. The model dependent quantities are presented in Section \ref{sec:different_models}. 

\subsection{Dynamical equations}
Consider a black hole (BH) with mass $M = 10^6 \cdot M_6$~$M_{\odot}$ and a star with mass $M_* = m_*$~$M_{\odot}$ and radius 
$R_* = x_*$~$R_{\odot}$. Here $M_6$ is the black hole mass in $10^6$ solar mass units, while $m_*$ and $x_*$ is the stellar mass and radius in solar mass units, respectively.

If a tidal disruption occurs, then the pericenter distance of the stellar orbit, $r_{\rm p}$, is similar to the tidal radius, $r_{\rm t}$, where tidal forces from the BH can completely disrupt the star \citep{1989IAUS..136..543P, 1989ApJ...346L..13E}: 
\begin{equation}
r_{\rm t} = R_* \cdot \left(\frac{M}{M_*}\right)^{1/3}  \simeq 0.47 x_* \cdot \left(\frac{M_6}{m_*}\right)^{1/3}  ~ \mathrm{AU}.
\end{equation}
Following LR11, we define the penetration factor $\beta$ as $\beta = r_{\rm t} / r_{\rm p}$. 
We note that many studies showed that $\beta = 1$ is not enough for the full disruption (e.g. GR13). We will discuss this further in Section \ref{sec:tmin}.

After complete disruption, almost half of the stellar debris remains bound to the BH and moves on an elliptical orbit. After $t_{\rm min}$ time the first debris particles return to the pericenter. The timescale for this phenomenon (also called as the fallback timescale $t_{\rm fb}$, \citealt{2021ARA&A..59...21G}) is the orbital period of the debris particles that were closest to the BH at the moment of tidal distance passage. We assume that the circularization timescale for the fallback material is very short, as it is seen in recent hydrodynamical simulations \citep{2022MNRAS.510.1627A, 2022arXiv220610641S}, even though it may not always be the case \citep{2015ApJ...804...85S}.
Thus, in our model $t_{\rm min}$ is practically the elapsed time between the moment of disruption and the start of the mass accretion onto the BH.

The bound stellar debris accretes onto the BH with an accretion rate $\dot{M}_{\rm fb}$, which depends on the distribution of the specific gravitational binding energy of the stellar debris 
($\epsilon$) .

If the accretion rate exceeds the Eddington rate, $\dot{M}_{\rm Edd} = L_{\rm Edd} / \eta c^2$, where $\eta$ is the efficiency of the conversion of kinetic energy into radiation, $L_{\rm Edd} = 4 \pi G c M / \kappa$ is the Eddington-luminosity of the BH mass ($\kappa$ is the opacity of the fallback material), then the accretion becomes temporarily super-Eddington. 
The photon pressure by the super-Eddington accretion luminosity expels a fraction of the fallback debris material, $M_{\rm out} = f_{\rm out} M_{\rm fb}$, in a super-Eddington wind, while 
$(1-f_{\rm out})$ part of the infalling debris quickly forms an accretion disk around the BH
\citep[e.g.][]{Strubbe09, Dai18, Curd19}.

\subsection{The light curve}
\label{sec:2.2:light_curve}

In the model outlined above the observable radiation from a TDE starts with a relatively fast initial peak caused by the super-Eddington wind, which is then followed by a slower luminosity variation due to the time-dependent accretion from the accretion disk. Note that this is true only for $M \lesssim 3 \times 10^7$~ $M_\odot$ BH masses \citep{2021ARA&A..59...21G}, because above this mass limit the accretion will likely remain sub-Eddington. Thus, the model we consider here is valid only below this BH mass limit. We also take into account the reprocessing of the high-energy photons from the disk to optical photons by the wind material \citep{Guillochon14, 2016MNRAS.455..859S}.

We describe the semi-analytic calculation of these two parts of the TDE light curve separately. The total luminosity is the sum of the wind and the disk contribution.  
Note that we do not consider the colliding streams scenario \citep[e.g.][]{2020MNRAS.492..686L} that predicts collisions of different streams from  the fallback stellar material producing a collision-induced outflow (CIO), which is a viable alternative for explaining the early-phase UV/optical peak. We also do not include the proposed interaction of the super-Eddington wind with the dusty torus around the BH in Active Galactic Nuclei (AGN) \citep{2021JHEAp..32...11Z}. The addition of these phenomena to the model is left for a future study.

Let us consider the super-Eddington wind first, since this component may be responsible for the bright, initial transient that may be visible for a shorter period of time ($\sim$ months). 

Based on SQ09, we adopt the following assumptions: $i)$ the kinetic energy of the fallback material is thermalized at 
\begin{equation}
    r_{\rm L} = \alpha R_{\rm S}
    \label{eq:rL_distance}
\end{equation}
distance from the BH ($\alpha > 1$) where $R_{\rm S}$ is the Schwarzschild radius, and $ii)$ the wind is launched from that distance with a velocity that is equal to the escape velocity at $r_{\rm L}$ multiplied by a constant parameter $f_{\rm v}$: 

\begin{equation}
    v_{\rm wind} ~=~ f_{\rm v} \cdot \sqrt{\frac{2 GM}{r_{\rm L}}}. 
    \label{eq:vwind}
\end{equation}

The energy rate deposited at $r_{\rm L}$ can be approximated as
\begin{equation}
    \dot{E}(r_{\rm L}) \approx 4 \pi r_{\rm L}^2 \sigma_{\rm SB} T_{\rm L}^4 = \frac{G M \dot{M}_{\rm fb}}{r_{\rm L}} = \eta \dot{M}_{\rm fb}c^2,
\label{eq:Econs}    
\end{equation}
where $\sigma_{\rm SB}$ is the Stefan-Boltzmann constant,  $T_{\rm L}$ is the temperature at $r_{\rm L}$ and $\eta = 1 / (2 \alpha)$. 

The temperature $T_{\rm L}$ can be derived from Equation~(\ref{eq:Econs}) as 
\begin{equation}
    \sigma_{\rm SB} T_{\rm L}^4 = \frac{c^6}{4 \pi G^2} \frac{\eta^3 \dot{M}_{\rm fb}}{M^2}.
    \label{eq:TL}
\end{equation}

As a first approximation, the radius of the photosphere, $r_{\rm ph}$, where the wind turns into being optically thin, can be estimated from the criterion 
$r_{\rm ph} \cdot \rho_{\rm ph} \cdot \kappa \approx 1$, where $\rho_{\rm ph}$ is the density at $r_{\rm ph}$ and $\kappa$ is the mean opacity. However, since the opacity is usually assumed to be dominated by electron scattering, photons are thermalized at a different depth called the color radius ($r_{\rm c}$). We examine the relation between $r_{\rm c}$ and $r_{\rm ph}$ in Section~\ref{sec:colorradius}. 

Assuming that the geometry of the outflow is spherical, the wind density at $r > r_{\rm L}$ can be calculated as
\begin{equation}
    \rho_w(r) = \frac{f_{\rm out}}{4 \pi r^2} \cdot \frac{\dot{M}_{\rm fb}}{v_{\rm wind}}.
\end{equation}

From the previous equations the radius of the photosphere can be calculated as:
\begin{equation}
    r_{\rm ph} = \frac{\kappa \cdot f_{\rm out} \cdot \dot{M}_{\rm fb}}{ 4 \pi v_{\rm wind}}
    \label{eq:rpht}
\end{equation}

Following SQ09, we assume that the outflow between $r_{\rm ph}$ and $r_{\rm L}$ is adiabatic, therefore the temperature of the photosphere can be derived from
\begin{equation}
T_{\rm ph} = T_{\rm L} \cdot (r_{\rm ph}/r_{\rm L})^{-2/3} \cdot (f_{\rm out}/f_{\rm v})^{1/3}
\label{eq:tpht}
\end{equation}

When the accretion rate changes in time, $r_{\rm ph}$ and $T_{\rm ph}$ also change accordingly.
Using Equations~(\ref{eq:rpht}) and (\ref{eq:tpht}) the monochromatic luminosity of the super-Eddington wind can be calculated as
\begin{equation}
L_{\rm w}(\nu) = 4 \pi^2r_{\rm ph}^2 \cdot B_{\nu}(T_{\rm ph}), 
\label{eq:Lwind}
\end{equation}
assuming that the photosphere radiates as a blackbody and the color radius is below the photosphere ($r_{\rm c} < r_{\rm ph}$). See Section~\ref{sec:colorradius} for the opposite case. 

The second source of radiation is the accretion disk formed by the fallback material around the BH.
This component is dominant in two cases: at high radiation energies (in the UV/X-ray domain) or at late phases (long after the disruption event), when the contribution from the super-Eddington wind is negligible or has decreased already. Indeed, late-time observations of TDEs taken 5-10 years after peak showed the existence of far-UV radiation that can be explained only with the presence of an unobscured, viscously and thermally stable accretion disk \citep{vanVelzen19}. 

We set the inner edge of the disk ($R_{\rm in}$) equal to the last stable circular orbit (LSCO) around a nonrotating BH, i.e. $R_{\rm in} \approx 3 R_{\rm S} = 6 GM/c^2$, while the outer edge ($R_{\rm out}$) is assumed being close to the circularization radius, i.e. $R_{\rm out} \approx 2 r_{\rm p}$. We adopt the semi-analytic slim disk model presented by SQ09, which is based on the following assumptions: i) the viscous time is much shorter than the fallback timescale, i.e. the disk is circularized before the matter gets accreted onto the BH; ii) radiation pressure dominates over the gas pressure; and iii) the viscous stress is proportional to the radiation pressure. Under these conditions the disk is thick and advective during the super-Eddington phase, but it is thin and cooling via radiative diffusion after the accretion turns into sub-Eddington. Note that during the early phases the disk may be unstable, and quickly spreads out from the LSCO up to beyond the circularization radius \citep[e.g.][]{2016MNRAS.455..859S}. Such a spreading disk can produce a strongly time-dependent radiation. It is not included in the disk model presented here, but left for a future study instead. 

The temperature of the disk strongly varies between the inner and the outer edge. The monochromatic flux emitted by a two-sided ring being at $r$ distance from the BH and having thickness of $dr$, can be calculated as
\begin{equation}
dF_{\nu} = 2 \cdot 2\pi r dr \cdot \pi B_{\nu}(T),
\end{equation}
where $B_{\nu}(T)$ is the Planck function ($\nu$ is the radiation frequency, $T$ is the temperature at $r$). 

The temperature of the assumed steady-state slim disk at a specific $r$ can be inferred from the following expression (SQ09):
\[
{\sigma}_{\rm SB}T^4=\frac{3GM\dot{M}f}{8\pi r^3} \times
\]
\begin{equation}
    \times \left[\frac{1}{2}+\left\{\frac{1}{4}+\frac{3}{2}\cdot f \cdot\left(\frac{\dot{M}}{\eta \dot{M}_{\rm Edd}}\right)^2 \left(\frac{r}{R_{\rm S}}\right)^{-2}\right\}^{1/2}\right]^{-1},
\end{equation}
where $\dot{M} = (1-f_{\rm out}) \dot{M}_{\rm fb}$ is the disk accretion rate, 
$f = 1 - \sqrt{R_{\rm in}/r}$, and $R_{\rm S} = 2GM/c^2$ is the Schwarzshild radius of the BH. This equation is slightly different from Equation~(27) in LR11, because we use Equation~(19) from SQ09.

Summing up the contribution from such rings between $R_{\rm in}$ and $R_{\rm out}$ gives the total monochromatic luminosity from the disk: 
\begin{equation}
    L_{\rm d}(\nu) ~=~ 2 \cdot \int_{R_{\rm in}}^{R_{\rm out}}  2 \pi r \cdot \pi B_\nu(T(r))~ dr
\end{equation}

Instead of simply adding $L_{\rm w}$ and $L_{\rm d}$ together, we also consider the reprocessing of the disk luminosity into optical radiation by the wind material \citep{Guillochon14, 2016MNRAS.455..859S}. We assume that a fraction of the bolometric luminosity of the disk is absorbed by the optically thick wind material, and after thermalization it increases the temperature of the wind at the photosphere: 
\begin{equation}
T_{\rm ph,rep}^4 = T_{\rm ph}^4 + \frac{\epsilon_{\rm rep} L_{\rm d}^{\rm bol} \cdot (1 - e^{-\tau_{\rm w}})}{4 \pi r_{\rm ph}^2 \sigma_{\rm SB}} ,
\label{eq:reproc}
\end{equation}
where $L_{\rm d}^{bol} = \int_0^\infty L_{\rm d}(\nu) d\nu$ is the total bolometric disk luminosity, $\epsilon_{\rm rep}$ is the efficiency of reprocessing, and $\tau_{\rm w}$ is the total optical depth of the wind for high-energy photons. We estimate $\tau_{\rm w}$ by assuming electron scattering (Thomson) opacity as 
\begin{equation}
    \tau_{\rm w} \simeq \int_{r_{\rm L}}^\infty \kappa \rho_{\rm w}(r) dr = \frac{\kappa f_{\rm out} \dot{M}_{\rm fb}}{4 \pi r_{\rm L} v_{\rm wind}}. 
    \label{eq:tw}
\end{equation}

Thus, Equation~(\ref{eq:tpht}) is supplemented by Equation~(\ref{eq:reproc}) and (\ref{eq:tw}) if reprocessing is non-negligible.

\section{Models for the key parameters}
\label{sec:different_models}
In this section we present some of the key parameters that have critical effect on the light curve of a TDE. These are the fallback timescale ($t_{\rm min}$), the time-dependent mass accretion rate ($\dot{M}_{\rm fb}$) and its peak value ($\dot{M}_{\rm p}$).  In the following we describe the various models for these parameters that are available in TiDE.

\subsection{The fallback accretion rate}
\label{sec:mdotfb_models}
The classical form of the fallback accretion rate ($\dot{M}_{\rm fb}$) assumes that $dM_*/d\epsilon$ is constant \citep{1988Natur.333..523R}. This is frequently called as the ``frozen-in'' model, which results in
\begin{equation}
    \dot{M}_{\rm fb}=\dot{M}_{\rm p}\left(\frac{t}{t_{\rm min}}\right)^{-5/3},
    \label{eq:mdotfb_classic}
\end{equation}
where $t_{\rm min}$ is the fallback timescale and $\dot{M}_p$ is the peak of the accretion rate. In the classical case it can be expressed as
\begin{equation}
    \dot{M}_{\rm p}=\frac{1}{3}\frac{M_{*}}{t_{\rm min}}
    \label{eq:mdotp_clasic}
\end{equation}
From Equation (\ref{eq:mdotfb_classic}) it follows that the accretion rate always decreases with time, since this model must be computed from $t = t_{\rm min}$. Thus, the peak of the accretion occurs at $t_{\rm min}$ ($t_{\rm peak} = t_{\rm min}$).

A more realistic model was presented by L09, which was used in many subsequent papers \citep[e.g.][]{Gallegos-Garcia18, Golightly_2, Golightly_1}. Its accretion rate can be expressed as
\begin{equation}
    \dot{M}_{\rm fb} = \frac{4 \pi}{3} \frac{R_*}{t_{\rm min}} \left( \frac{t}{t_{\rm min}} \right)^{-5/3} \int\displaylimits_{x(t)}^{R_*} \rho(R) R dR, 
    \label{eq:mdotfb_l09_all}
\end{equation}
where $x(t) = R_* (t/t_{\rm min})^{-2/3}$ and $\rho(R)$ is the density profile of the star before the disruption occurs.
In TiDE we solve a polytropic stellar model with two different polytropic index: n=3 ($\gamma = 4/3$) and n=3/2 ($\gamma = 5/3$). We apply two different mass-radius relations: $x_* = m_*^\delta$ for main sequence stars with $\delta = 0.8$ if $m_* < 1$ and $\delta = 0.5$ if $m_* > 1$ \citep{kw94}, and $x_* = 0.01 m_* ^{-1/3}$ for white dwarfs. The moment of the peak accretion rate can be calculated numerically from the time derivative of Equation (\ref{eq:mdotfb_l09_all}): it is $t_{\rm peak} = 5.77 \cdot t_{\rm min}$ and $t_{\rm peak} = 2.66 \cdot t_{\rm min}$ for $n = 3$ and $n = 3/2$, respectively.

For comparison, we also use the special case when $\rho(R)$ is constant. In this case the resulting accretion rate is
\begin{equation}
    \dot{M}_{\rm fb}(t) = \dot{M_{\rm p}} \left [ 1 - \left (\frac{t}{t_{\rm min}} \right)^{-4/3}  \right ] \cdot \left ( \frac{t}{t_{\rm min}}  \right )^{-5/3},
    \label{eq:mdfb_lod09_const}
\end{equation}
where
\begin{equation}
    \dot{M}_{\rm p} = \frac{1}{2} \frac{M_*}{t_{\rm min}}
\end{equation}
In this approximation, the peak of the accretion rate occurs at $t_{\rm peak} = 1.55 \cdot t_{\rm min}$.

\begin{figure*}
    \centering
    \includegraphics[width=1.5\columnwidth]{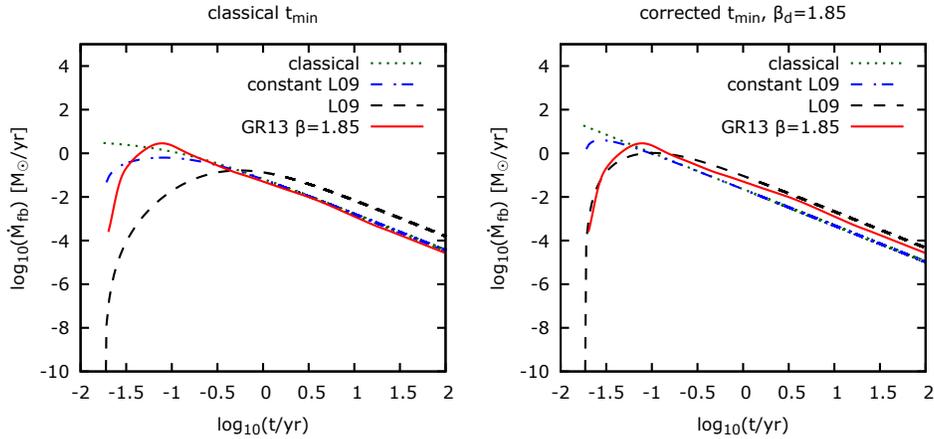}
    \caption{Different $\dot{M}_{\rm fb}$ calculations (dotted green: classical $\dot{M}_{\rm fb}$, dash-dotted blue: L09 accretion rate with constant $\rho$ approximation, dashed black: L09 accretion rate, continuous red: GR13 hydrodynamical simulation). 
    The left panel represents the accretion rates that 
    were calculated with the classical $t_{\rm min}$ (Eq. \ref{eq:tmin_classical}). Note that all curves were shifted horizontally to get the same starting point as the GR13 curve. The curves in the right panel are calculated with the corrected $t_{\rm min}$ and with $\beta_{\rm d} = 1.85$ (see Eq. \ref{eq:tmin_corrected}). In this case there is no need for any horizontal shift. The applied parameters are presented at Table \ref{Table:initial_parameters}. Also, $\gamma = 4/3$ was assumed.}
    \label{fig:mdotfb_comparsion_4per3}
\end{figure*}

\begin{figure*}
    \centering
    \includegraphics[width=1.5\columnwidth]{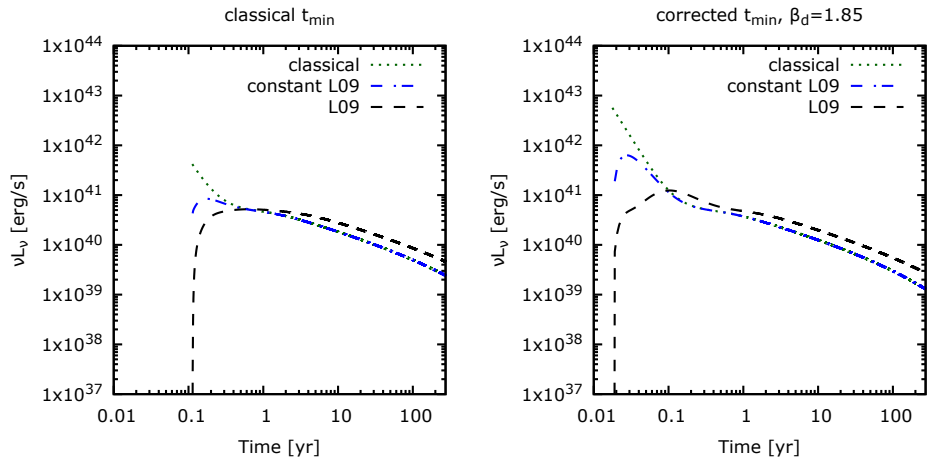}
    \caption{The calculated light curves of different fallback accretion models in optical g band. The notation and the used parameters are the same as in Figure \ref{fig:mdotfb_comparsion_4per3}. Note that no horizontal shift was applied in the classical $t_{\rm min}$ case (left panel).}
    \label{fig:lcurves_different_mdotfb}
\end{figure*}

The differences between these model accretion rates can be seen in Figure~\ref{fig:mdotfb_comparsion_4per3}, while the corresponding light curves are plotted in Figure \ref{fig:lcurves_different_mdotfb}.

\begin{table}
    \centering
    \caption{Parameters for the fiducial model}
    \begin{tabular}{ccccccc}
    \hline
    \hline
    $M_6$ & $m_*$ & $x_*$ & $\eta$ & $\beta$ & $f_{\rm out}$ & $f_{\rm v}$\\ 
    \hline
    1.0 & 1.0 & 1.0 & 0.1 & 1.0 & 0.1 & 1.0 \\
    \hline
    \end{tabular}
    \label{Table:initial_parameters}
\end{table}

\subsection{Parameters from simulations by GR13}
\label{sec:GR13_parameters}

In GR13 the authors computed many hydrodynamical simulations modeling the disruptions of polytropic stars, and investigated the fallback accretion rates. They presented analytical fits to the peak of the accretion rate and the time of the light curve peak as a function of $M_6$, $m_*$, $x_*$ and $\beta$ (see their A1-A2 equations).
These results and the exact fallback curves are useful for testing the results of the different models adopted in TiDE.

\subsection{$t_{\rm min}$ definitions}
\label{sec:tmin}
The classical value of $t_{\rm min}$ is the orbital period of the particle that is closest to the BH when the star reaches the tidal radius \citep[e.g.][]{2013MNRAS.435.1809S}:
\begin{equation}
    t_{\rm min} = \frac{\pi}{\sqrt{2}}\left(\frac{r_{\rm t}}{R_{*}}\right)^{3/2} \sqrt{\frac{r_{\rm t}^3}{GM}}
    \label{eq:tmin_classical}
\end{equation}

Many previous studies \citep[for example][]{Lodato11, Strubbe09} adopted a similar, but not the same equation to determine $t_{\rm min}$; they used the pericenter distance, $r_{\rm p}$, instead of $r_{\rm t}$:
\begin{equation}
    t_{\rm min} = \frac{\pi}{\sqrt{2}}\left(\frac{r_{\rm p}}{R_{*}}\right)^{3/2} \sqrt{\frac{r_{\rm p}^3}{GM}}
    \label{eq:tmin_with_rp}
\end{equation}

Here we introduce another $t_{\rm min}$ definition based on the following argument: many studies showed that the full disruption does not occur when $\beta = 1$  (e.g. GR13; \citealt{Golightly_2}). It follows that simply reaching the classical $r_{\rm t}$ distance is not enough for the total disintegration. But the classical $t_{\rm min}$ was definied by using the classical $r_{\rm t}$ distance. Here we introduce a modified $r_{\rm t}$ parameter, namely
\begin{equation}
    r'_{\rm t} = \beta' r_{\rm p}
\end{equation}
and assume that full disruption happens when the star is able to reach $\beta' = 1$ after passing the tidal radius.
With the initial notation, this happens at $\beta = \beta_{\rm d} > 1$.
Therefore, $r'_{\rm t} = r_{\rm t}/\beta_{\rm d}$ is the distance from the BH where the star is fully disrupted. Using this definition our corrected $t_{\rm min}$ becomes
\begin{equation}
    t_{\rm min} = \frac{\pi}{\sqrt{2}}\left(\frac{r'_{\rm t}}{R_{*}}\right)^{3/2} \sqrt{\frac{(r'_{\rm t})^3}{GM}}
    \label{eq:tmin_corrected}
\end{equation}
If $\beta_{\rm d} = 1$ it restores the classical definition (Eq. \ref{eq:tmin_classical}). 

We also adopt yet another method for calculating $t_{\rm min}$ based on the fitted $t_{\rm peak}$ values of GR13 (hereafter: $t_{\rm peak,GR13}$). We know that the peak time of the accretion rate depends linearly on $t_{\rm min}$: $t_{\rm peak} = C \cdot t_{\rm min}$, where the $C$ factor is actually a function of the polytropic index as discussed in Section \ref{sec:mdotfb_models}. If we use the $t_{\rm peak,GR13}$ values, $t_{\rm min}$ can be calculated as
{
\begin{equation}
    t_{\rm min} = \frac{t_{\rm peak,GR13}}{C}
    \label{eq:tmin_from_GR13}
\end{equation}
}
In this case, $t_{\rm min}$ and therefore the light curve will be more dependent on $\beta$.

Here we introduce the fallback accretion rates and their light curves with different $t_{\rm min}$ definitions. In Figure \ref{fig:mdotfb_comparsion_4per3} one can see the fallback accretion rate computed with the classical $t_{\rm min}$ (left panel) and with the corrected $t_{\rm min}$ assuming $\beta_{\rm d} = 1.85$ where the full disruption of a $\gamma=4/3$ star happens according to GR13 (right panel), both compared to the accretions rates from hydrodynamical simulations by GR13. In the left panel neither the time of the peak nor the maximum accretion rate match with the values given by GR13, thus, a horizontal shift was needed to bring the different curves to the same starting point. This is not a problem with the model of the L09 accretion rate (right panel, black, dashed line) as it matches well with the GR13 value. 

In Figure \ref{fig:lcurves_different_mdotfb} the light curves in the optical $g$-band are shown. It is seen that the peak luminosity values from the various models differ by about an order of magnitude.

\begin{figure*}
    \centering
    \includegraphics[width=1.5\columnwidth]{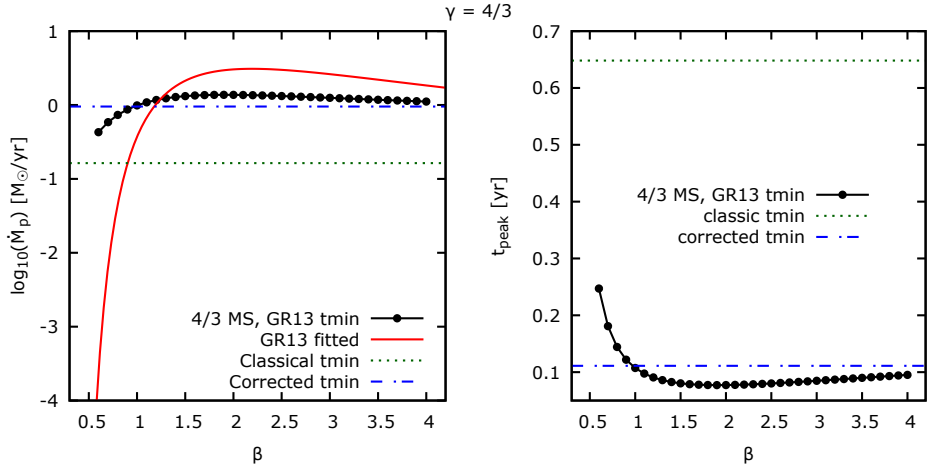}
    \caption{The peak of the accretion rate $\dot{M}_{\rm p}$ (left panel) and its timescale $t_{\rm peak}$ (right panel) as a function of $\beta$ for the different $t_{\rm min}$ models with the L09 accretion rate and $\gamma=4/3$. The red solid curve is the peak accretion rate function given by GR13. Black curve with circles is the calculated $\dot{M}_{\rm p}$ given by Eq.~\ref{eq:tmin_from_GR13}. This gives the same $t_{\rm peak}$ values as in GR13, hence the latter is not plotted in the right panel. Green dotted lines denote the $\dot{M}_{\rm p}$ and $t_{\rm peak}$ values corresponding to the classical $t_{\rm min}$ formula (Eq.~\ref{eq:tmin_classical}). Blue dashed-dotted line represents the calculated values of the corrected form with $\beta_d=1.85$ (Eq.~\ref{eq:tmin_corrected}). It is seen that the classical $t_{\rm min}$ definition cannot predict either the accretion rate or the peak time, while both the corrected and the GR13-based $t_{\rm min}$ calculations are in fairly good agreement with the SPH simulations of GR13 in the case of full disruptions ($\beta >1$), especially around $\beta\approx 1.2$.  
    }
    \label{fig:mdp_tpeak_4per3}
\end{figure*}

In Figure \ref{fig:mdp_tpeak_4per3} we investigate the time and the value of the peak of the accretion rate as a function of $\beta$ compared to the values given by GR13 for a main sequence star with $\gamma = 4/3$. The accretion model is L09 in this case, thus,  $t_{\rm peak} = 5.77 \cdot t_{\rm min}$. Here we compare the predictions from the three different $t_{\rm min}$ definitions outlined above: the classical model (Equation~\ref{eq:tmin_classical}; green dotted line), the corrected $t_{\rm min}$ model (Equation~\ref{eq:tmin_corrected}; blue dash-dotted line) and finally the model that calculates $t_{\rm min}$ directly from the $t_{\rm peak,GR13}$ values 

(Equation~\ref{eq:tmin_from_GR13}; black symbols).  The left panel shows the predictions for the peak of the accretion rate ($\dot{M}_{\rm p}$). 
It is seen that the curve computed with the corrected $t_{\rm min}$ (Equation~\ref{eq:tmin_corrected}; blue dash-dotted line) fits the full disruption part of the GR13 results (shown by the red curve) better than the one using the classical $t_{\rm min}$ definition (green dotted line), at least for $\beta > 1$ where full disruption is likely. There is also a reasonable agreement with the predictions using Equation~\ref{eq:tmin_from_GR13} (black curve with dots), again, for $\beta > 1$. It is also clear that none of these $t_{\rm min}$ definitions work for partial disruptions ($\beta \lesssim 1$). The right panel of Figure~\ref{fig:mdp_tpeak_4per3} displays the same comparison, but for the $t_{\rm peak}$ parameter.
In this case we do not plot the GR13 fitted values (red curve in the left panel), because the $t_{\rm peak}$ values plotted with black symbols are the same as the $t_{\rm peak,GR13}$ values.

\begin{figure*}
    \centering
    \includegraphics[width=1.5\columnwidth]{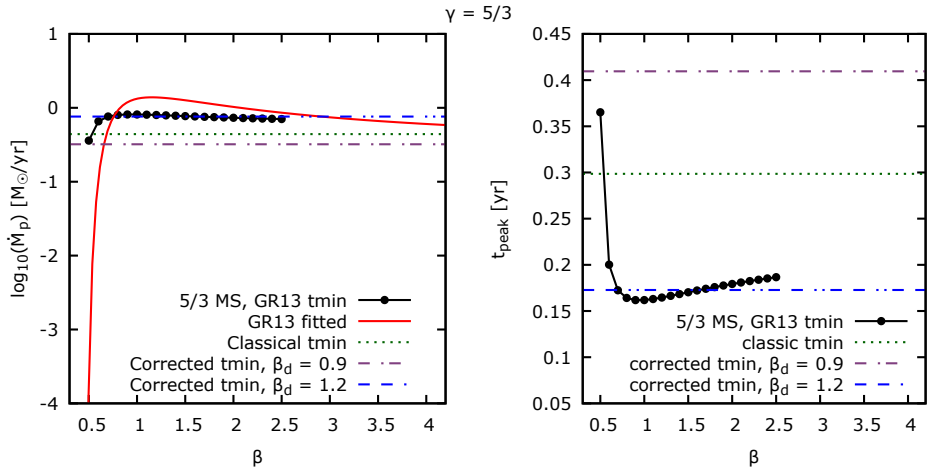}
    \caption{Same as Figure \ref{fig:mdp_tpeak_4per3} but for $\gamma = 5/3$. Here we use two different $\beta_d$ parameter: $\beta_d=0.9$ is the full disruption limit in GR13 (purple dash-dotted line) and $\beta_d=1.2$ (blue double-dash-double-dotted line) which brings the original GR13 fits (red curve) and the GR13-based values (black filled circles) into agreement.}
    \label{fig:mdp_tpeak_5per3}
\end{figure*}

Figure \ref{fig:mdp_tpeak_5per3} is the same as Figure~\ref{fig:mdp_tpeak_4per3} but it shows the results for the disruptions of a $\gamma = 5/3$ star. In this case it is found that the corrected $t_{\rm min}$ definition does not work well if we use $\beta_{\rm d}=0.9$ that corresponds to the full disruption in GR13. In the $\gamma = 5/3$ case it is better to use $\beta_{\rm d} = 1.2$ while estimating $t_{\rm min}$ from Equation~\ref{eq:tmin_corrected}.

\begin{figure}
    \centering
    \includegraphics[width=\columnwidth]{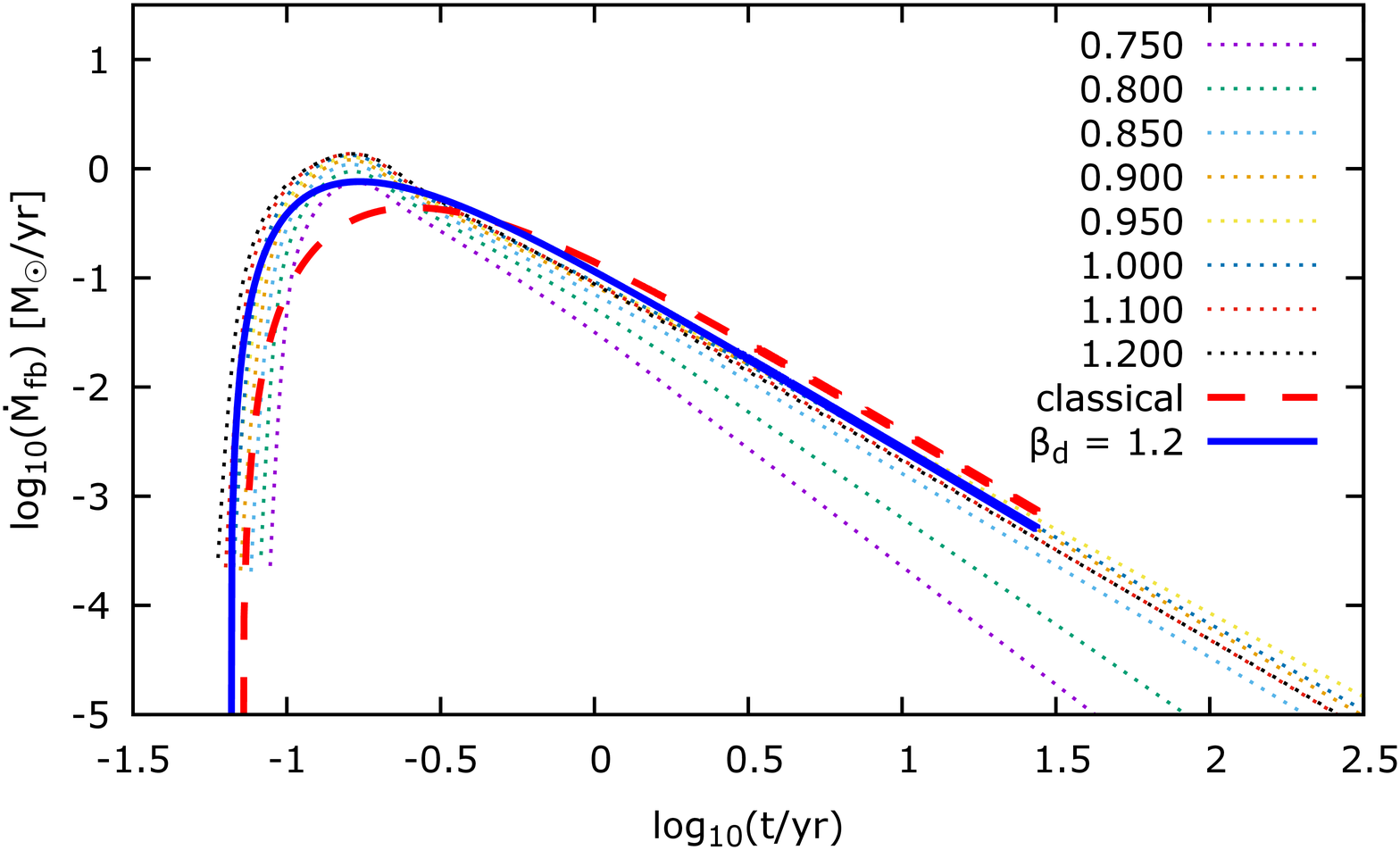}
    \caption{The accretion rates with different models of a $\gamma = 5/3$ star. Dotted curves are the simulations by GR13 for different values of $\beta$, the red dashed curve is calcuated by TiDE using the classical $t_{\rm min}$ model (shifted horizontally to match the moment of the start of the accretion in the simulations). The blue continuous curve is the TiDE model using the corrected $t_{\rm min}$ definition with $\beta_{\rm d} = 1.2$, which fits the GR13 simulations much better than the model computed from the classical $t_{\rm min}$. The agreement gets worse for $\beta < 1$, i.e. for partial disruptions (see text).}
    \label{fig:mdotfb_comparsion_5per3}
\end{figure}

Figure ~\ref{fig:mdotfb_comparsion_5per3} is similar to Figure~2 in \citet{Gallegos-Garcia18}, but we also present our corrected $t_{\rm min}$ curve with $\beta_{\rm d} = 1.2$ (blue continuous line). For the classical $t_{\rm min}$ curve (red dashed line) a horizontal shift was applied to match the moment of the start of the accretion with that of the GR13 simulation. Again, the rates derived from the corrected $t_{\rm min}$ definition provide a much better fit to the GR13 results, and in this case the horizontal shift is not necessary.

\subsection{Time dependent $f_{\rm out}$ parameter}
\label{sec:fout}

The original TDE model outlined in Section~\ref{sec:tde_model} contains an $f_{\rm out}$ parameter that connects the mass of the outflow material expelled during the super-Eddington phase to the infalling debris mass. In the simplest case $f_{\rm out}$ is a constant and independent from any other parameters. We may relax this constraint by adopting Equation~(28) from LR11 as
\begin{equation}
f_{\rm out} = \frac{2}{\pi}\mathrm{arctan}\left[\frac{1}{7.5}\left(\frac{\dot{M}_{\rm fb}}{\dot{M}_{\rm Edd}} - 1 \right)\right].
    \label{eq:timedependfout}
\end{equation}

This formula is based on the work by \citet{2011MNRAS.413.1623D}. 
Because $\dot{M}_{\rm fb}$ is time dependent, this new $f_{\rm out}$ will also be time dependent. 

LR11 note that for $f_{\rm out} \gtrsim 0.7$ (about $\dot{M}_{\rm fb}/\dot{M}_{\rm Edd} \gtrsim 20$), Equation~(\ref{eq:timedependfout}) may not be correct. Also, it can be applied only in the case of super-Eddington accretion, i.e.
when $\dot{M}_{\rm fb} > \dot{M}_{\rm Edd}$. 

\begin{figure}
    \centering
    \includegraphics[width=\columnwidth]{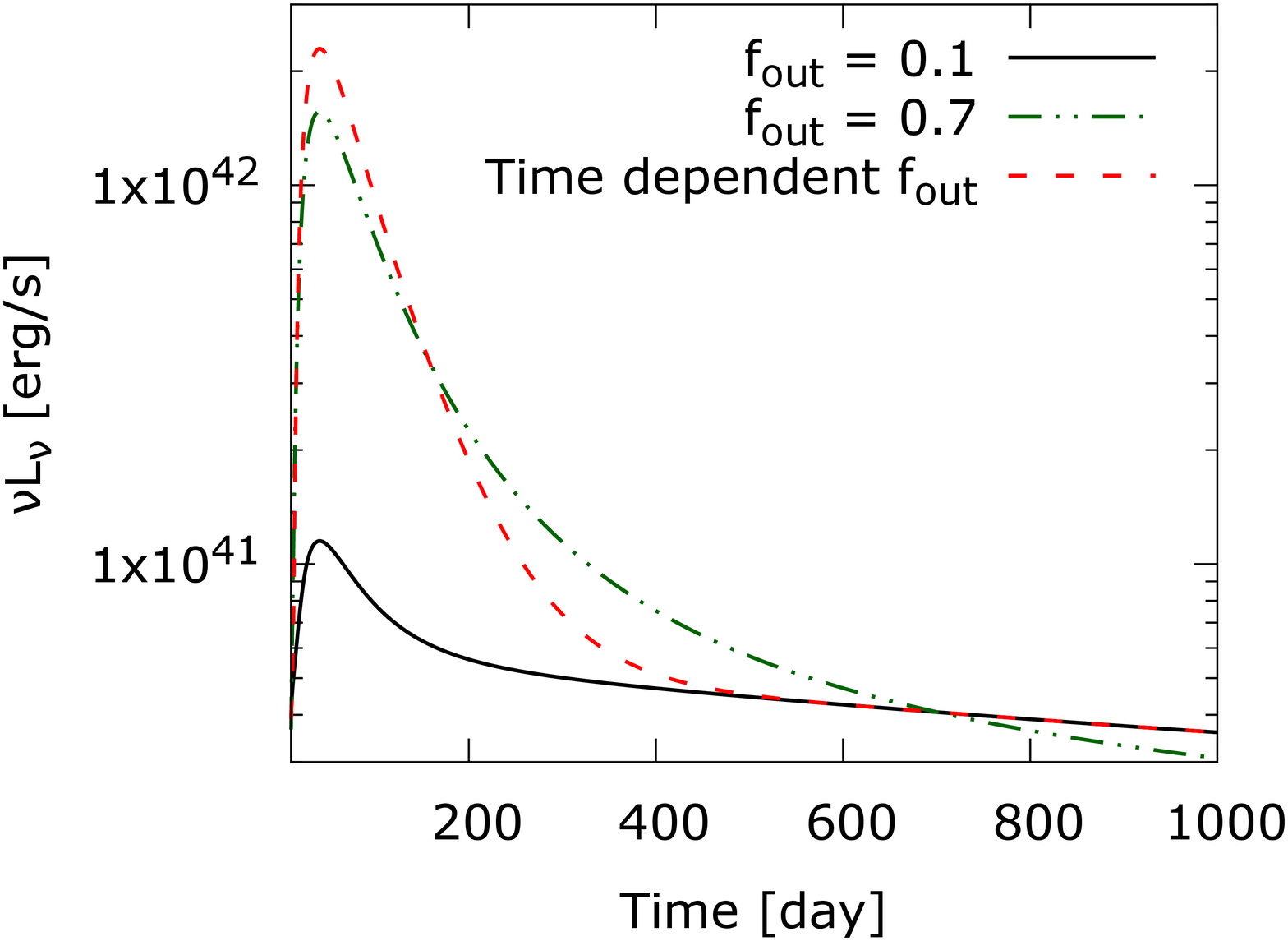}
    \caption{The effect of the time-dependent $f_{\rm out}$ parameter on the light curve. All other parameters are the fiducial ones (see Table \ref{Table:initial_parameters}), except that $\eta = 0.02$ has been used to produce this figure. Black and green lines represent the light curves with constant $f_{\rm out}$ (black: $0.1$, green: $0.7$) while the red line is the light curve computed with the time-dependent $f_{\rm out}$ parameter.}
    \label{fig:timedependent_fout}
\end{figure}

The effect of the time-dependent $f_{\rm out}$ parameter on the fiducial light curve can be seen in Figure~\ref{fig:timedependent_fout}. We also show the calculated light curves with constant $f_{\rm out} = 0.1$ (black line) and $0.7$ (green dash-dotted line), while all the other parameters are the same as in Table~\ref{Table:initial_parameters}. It is seen that the time-dependent $f_{\rm out}$ parameter results in higher luminosity (red curve) than that of the fiducial model (black curve), and also different from the $f_{\rm out} = 0.7$ case (green curve). The decline rates of this time-dependent case are also very different from the constant case.

\subsection{Photospheric, trapping and color radius}
\label{sec:colorradius}

In Section \ref{sec:2.2:light_curve} we used the photospheric radius, $r_{\rm ph}$, to calculate the luminosity of the super-Eddington wind component. We adopted the definition of $r_{\rm ph}$ as $\kappa \rho(r_{\rm ph}) r_{\rm ph} = 1$, where $\kappa$ is the mean opacity, which is composed of $\kappa_{\rm es}$ electron scattering and $\kappa{\rm a}$ absorption opacities. Since $\kappa_{\rm es} >> \kappa{\rm a}$ \citep{Matsumoto}, $\kappa \simeq \kappa_{\rm es}$.

The trapping radius ($r_{\rm tr}$, also called as the diffusion radius $R_{\rm d}$ in \citet{Matsumoto}) is the distance where the photon diffusion time is equivalent with the dynamical time \citep{Piro}. Most papers use the criterion that at this distance
\begin{equation}
    \kappa \rho(r_{\rm tr}) r_{\rm tr} = \frac{c}{v_{\rm wind}}.
    \label{eq:rtr}
\end{equation}
\citet{Piro} showed that $r_{\rm tr}$ can be calculated as
\begin{equation}
    \frac{r_{\rm tr}}{r_{\rm L}} = 1 + \frac{\kappa_{\rm es} f_{\rm out} \dot{M}_{\rm fb}}{4 \pi r_{\rm L} c} \frac{(r_{\rm w} - r_{\rm tr})^2}{r_{\rm w}^2},
    \label{eq:rtr_full}
\end{equation}
where $r_{\rm w} = r_{\rm L} + (t-t_{\rm min}) v_{\rm w}$ is the outer boundary of the wind.

Equation~(\ref{eq:rtr_full}) simplifies into Equation~(\ref{eq:rtr}) only in certain cases (see \citet{Piro} for more discussion).

The color radius ($r_{\rm c}$) is the location where the last absorption of the photons occurs. In other papers, e.g. \citet{Shen}, this is called as the thermalization radius, $R_{\rm th}$. The value of $r_{\rm c}$ can be calculated using the following criterion: 
\begin{equation}
    \tau_{\rm eff} (r_{\rm c}) \equiv \int\displaylimits_{r_{\rm c}}^{\infty} \kappa_{\rm eff} \rho dr = 1,
    \label{eq:kappaeff}
\end{equation}
where $\kappa_{\rm eff} = (3 \kappa_{\rm es} \kappa_{\rm a})^{1/2}$ \citep{Piro}.

Here we approximate the absorption opacity $\kappa_{\rm a}$ with Kramer's opacity law:
\begin{equation}
    \kappa_{\rm a} = \kappa_0 \rho T^{-7/2}, 
    \label{eq:kramers}
\end{equation}
where $\kappa_0 \approx 4 \times 10^{25}$ in cgs units \citep{Matsumoto}.

In order to calculate the luminosity from the super-Eddington wind, the relation between $r_{\rm c}$ and $r_{\rm tr}$ must be determined. 

If $r_{\rm c} < r_{\rm tr}$, the wind luminosity can be derived from Equation~(\ref{eq:Lwind}) using $r_{\rm ph} \approx r_{\rm tr}$.
However, if $r_{\rm c} > r_{\rm tr}$, one needs to replace $r_{\rm ph}$ with $r_{\rm c}$ and $T_{\rm ph}$ with $T_{\rm c}$ in Equation~(\ref{eq:Lwind}). In this latter case the photons are adiabatically cooled to $r_{\rm tr}$, then keep constant luminosity up to $r_{\rm c}$ \citep{Piro}. 
Our TiDE code calculates both $r_{\rm c}$ and $r_{\rm tr}$, then uses the larger one for $r_{\rm ph}$ and determines the corresponding $T_{\rm ph}$. 
\citet{Piro} showed that for typical TDEs the color radius is usually smaller than the trapping radius. 
This is supported by our calculations as well, since for typical $M_6$, $m_*$ and  $\eta$ values we found that $r_{\rm c} < r_{\rm tr}$.

We also examine the possible conditions that can result in $r_{\rm c} > r_{\rm tr}$. We calculate the ratio between $r_{\rm c}$ and $r_{\rm tr}$ by combining Equation~(\ref{eq:rtr}), (\ref{eq:kappaeff}) and (\ref{eq:kramers}):

\begin{equation}
    \frac{r_{\rm c}}{r_{\rm tr}} =\frac{3^{4/9} \sigma^{7/18} \kappa_0^{4/9} f_{\rm out}^{2/27} \dot{M}_{\rm fb}^{11/54}}{2^{53/27} \pi^{11/54} \kappa_{\rm es}^{8/27} f_{\rm v}^{22/27} c^{23/27} (G M)^{7/27} \eta^{43/54}}.
    \label{eq:rcrpertr}
\end{equation}

This ratio is greater than 1 only if $\eta \lesssim 10^{-3}$ for typical $M_6$ masses, solar-like stars and L09 accretion rates.
Such a small $\eta$ is out of range of the possible $\eta$ values that are usually adopted in the literature, namely $0.04 \leq \eta \leq 0.42$ \citep{1969Natur.223..690L, 1970Natur.226...64B}. This is even smaller then the values found by MOSFiT e.g. in \citet{2019ApJ...872..151M}, or \citet{2020MNRAS.499..482N}, which makes this scenario very unlikely.

\subsection{Photon diffusion}

\label{sec:diffusion}

Shortly after the beginning of the fallback accretion, when the super-Eddington outflow starts, the debris close to the BH is presumably dense, i.e. optically thick. As a result, the accretion-generated radiation must go through this dense wind material via radiative diffusion \citep{2016MNRAS.461..948M}. In this case the observed luminosity is delayed with respect to the accretion luminosity by a characteristic timescale $t_{\rm diff}$. This process can be added to the light curve model by calculating the observed luminosity via the following equation:
\begin{equation}
    L(t^*) = \frac{1}{t_{\rm diff}} e^{-\frac{t^*}{t_{\rm diff}}} \int_{0}^{t^*} e^{\frac{t'}{t_{\rm diff}}} L_{\rm inp}(t') dt'
\label{eq:diffusion}
\end{equation}
where $t_{\rm diff}$ is the diffusion timescale, $t^* = t - t_{\rm min}$ i.e. the elapsed time since the arrival of the first fallback material, while $L_{\rm inp}(t)$ is the accretion luminosity. We assume that $L_{\rm inp}(t)$ is equal to the sum of the wind and the disk component.  

\begin{figure}
    \centering
    \includegraphics[width=\columnwidth]{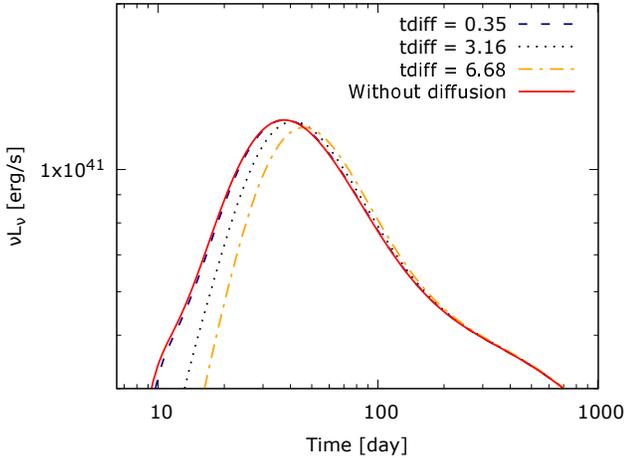}
    \caption{The light curves of a TDE with different diffusion timescales. All other parameters are the fiducial ones (Table \ref{Table:initial_parameters}).}
    \label{fig:diffusion_lcurves}
\end{figure}

Figure~\ref{fig:diffusion_lcurves} illustrates the effect of diffusion on the shape of the light curve. It is seen that a lower diffusion timescale results in higher maximum luminosity, but lower rise time, as expected.

Equation~(\ref{eq:diffusion}) is identical to modeling the build-up of the accretion disk due to viscous processes \citep{2019ApJ...872..151M}, which is implemented in the MOSFiT code. Note that in our interpretation the luminosity is spread out in time instead of the mass accretion rate, but otherwise the pre-maximum light curve shapes in these two approaches are similar.  

\section{The light curve of a tidal disruption event}
\label{sec:light_curves_comparsion}
    
\subsection{The effects of different parameters on the light curve}

\begin{figure*}
    \centering
    \includegraphics[width=1.7\columnwidth]{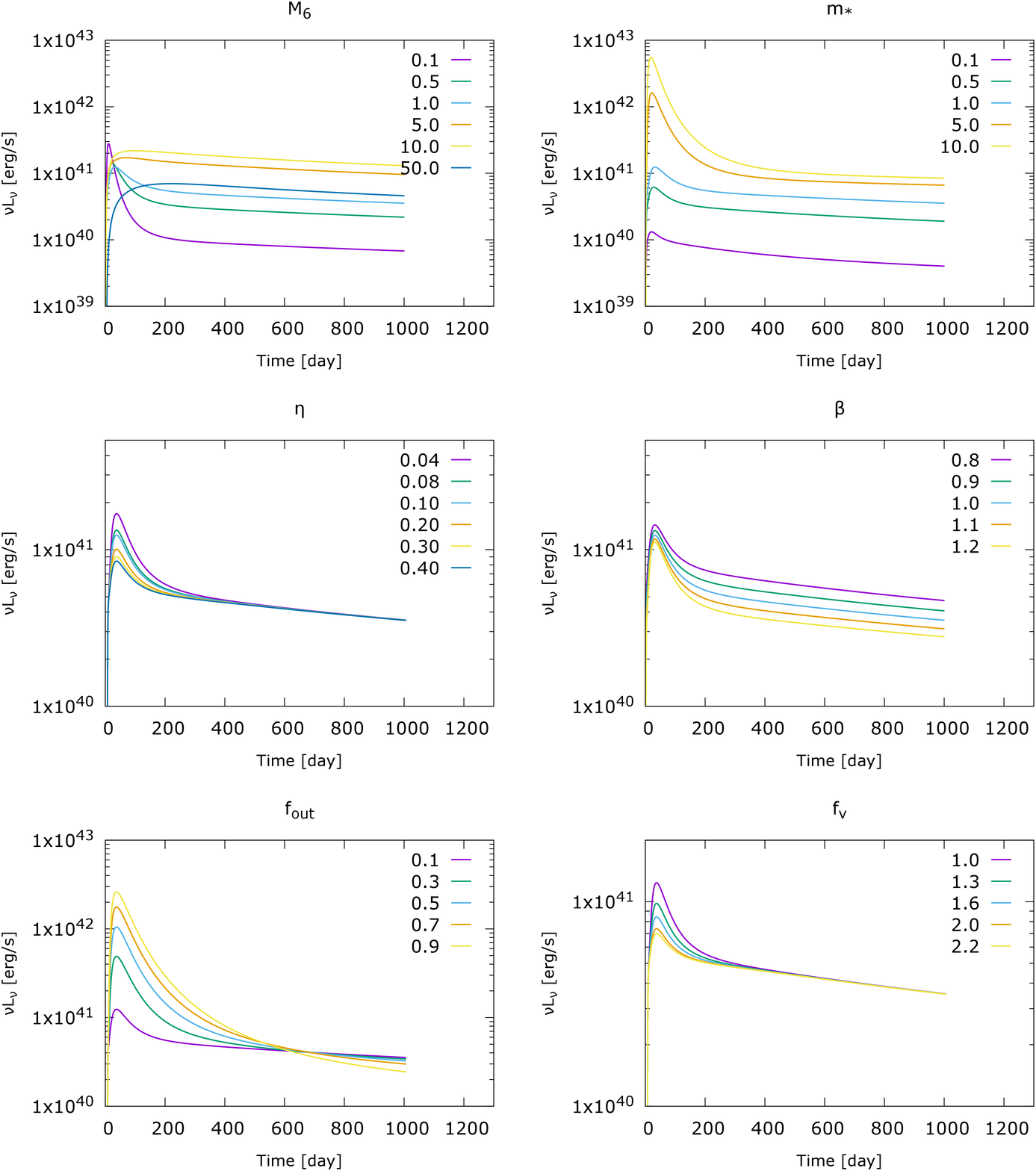}
    \caption{The effect of different parameters on the fiducial light curve. Each panel represents one changing parameter (indicated in the legend), while all the others are fixed at their initial values: $M_6 = 1$, $m_* = 1$, $x_* = 1$, $\eta = 0.1$, $\beta = 1$, $f_{\rm out} = 0.1$, $f_{\rm v} = 1$ (see Table~\ref{Table:initial_parameters})}
    \label{fig:parameter_change}
\end{figure*}

\begin{figure*}
    \centering
    \includegraphics[width=1.5\columnwidth]{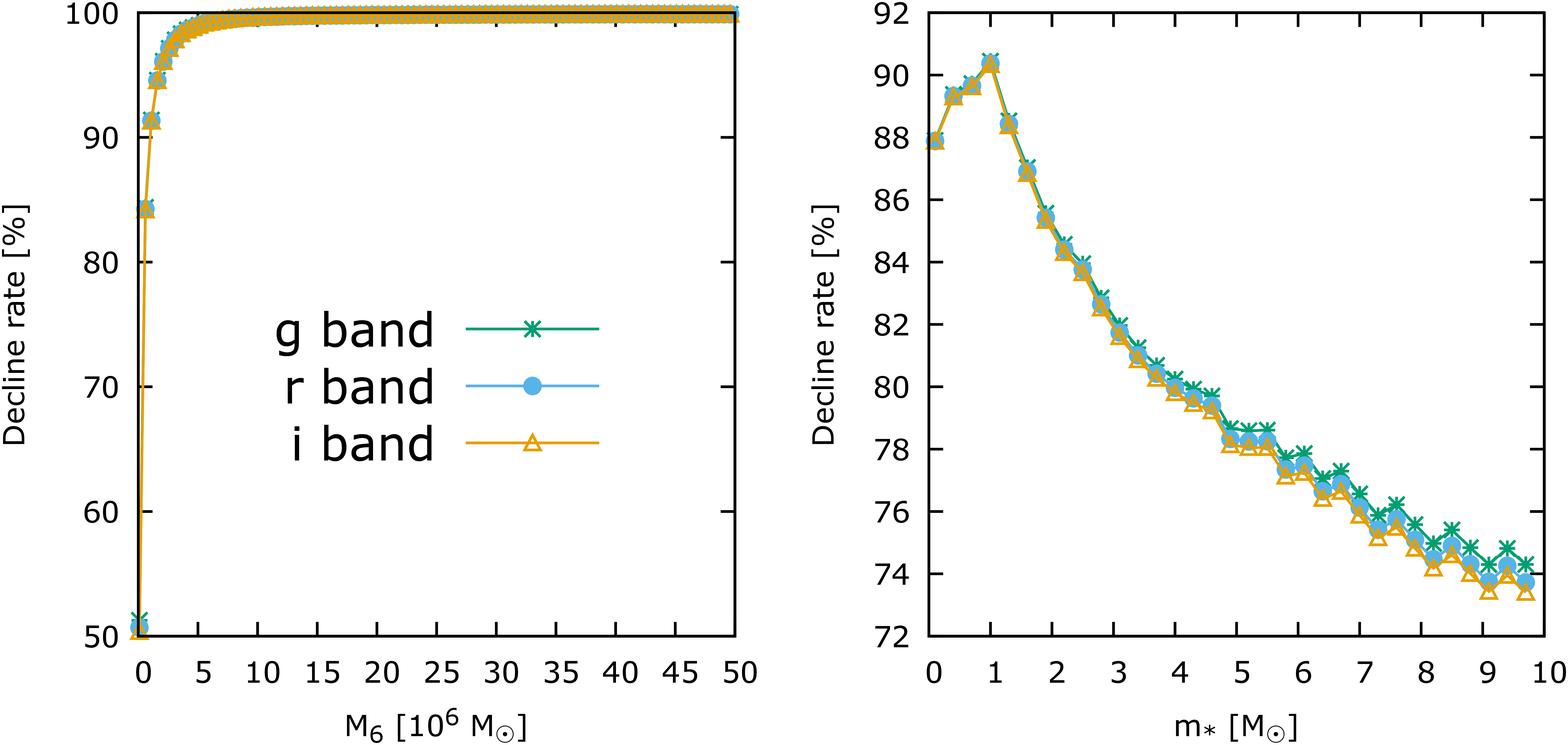}
    \caption{Decline rates measured at 15 days after maximums as functions of model parameters. Different colors represent different bands as indicated.}
    \label{fig:fading_rates_without_diffusion}
\end{figure*}

How the variation of the model parameters affects the
model light curve in the optical g band is seen in Figure \ref{fig:parameter_change}. All panels in this Figure represent the total (wind plus disk) luminosity of the TDE. In all cases we use the L09 accretion model (Equation~\ref{eq:mdotfb_l09_all}) with $\gamma = 4/3$ and our corrected $t_{\rm min}$ definition (Equation~\ref{eq:tmin_corrected}) with $\beta_{\rm d} = 1.8$. The light curves are shown at and after their corresponding $t_{\rm min}$ value, but for plotting purposes they are shifted horizontally to a common starting point in the panels showing the $M_6$ and $m_*$ dependence.

The top left panel in Figure \ref{fig:parameter_change} shows that increasing the BH mass ($M_6$) causes a decrease in the peak luminosity, while the light curve timescale increases.

The top right panel illustrates the effect of increasing stellar mass ($m_*$). The disruption of a more massive star results in a more luminous light curve. It is expected, since a more massive stellar debris means higher mass accretion rate, hence a higher luminosity.

The radiative efficiency parameter ($\eta$), however, does not cause any major variation either in the shape or the peak of the light curve (Figure \ref{fig:parameter_change} middle left panel) and the same is true for $\beta$ (middle right panel).

The $\eta$ dependency on the light curve first stems from Equation~(\ref{eq:rL_distance}). Previous papers (e.g. \citet{Strubbe09}, \citet{Lodato11}) adopted $r_{\rm L} = 2r_{\rm p}$ and from it this definition eliminated $\eta$ from their equations. In our model, if we use the time dependent $f_{\rm out}$ parameter (Equation~\ref{eq:timedependfout}), the final light curve will depend on $\eta$. 

Another interesting finding is that based on Figure \ref{fig:parameter_change} the peak of the light curve decreases with increasing $\eta$. This is because in Equation~(\ref{eq:Lwind}) $r_{\rm ph}^2 \sim \eta^{-1}$, while the temperature increases with $\eta$. So, the global outcome from the product of these two factors depends on the adopted wavelength where the blackbody function is evaluated. Thus, the intuition that higher radiative efficiency should result in a higher luminosity is true only for the bolometric light curve and at shorter (EUV) wavelengths.

In the middle right panel it is seen that the peak of the light curve is practically not affected by the $\beta$ parameter. This is expected because our adopted wind model is independent from $\beta$. However, there is a slight dependence on $\beta$ at late phases (after $\sim 200$ days) when the disk starts dominating the light curve. This is because the value of $\beta$ adjusts the outer edge of the accretion disk.

The $f_{\rm out}$ parameter is related to the mass of the wind in a sense that an increasing $f_{\rm out}$ means a more massive wind and a less massive disk component. The bottom left panel in Figure \ref{fig:parameter_change} shows that increasing $f_{\rm out}$ causes a more luminous and more slowly declining light curve. It is also an expected result, because during the early, wind-dominated phases a more massive wind results in a more luminous maximum and takes longer to decline.

From Equation~(\ref{eq:vwind}) we have an upper limit for the $f_{\rm v}$ parameter, because the wind velocity cannot exceed the speed of light. From this constraint the maximum value is $f_{\rm v} = 2.23$  in the fiducial model. From the lower right panel of Figure~\ref{fig:parameter_change} it is seen that the increase of $f_{\rm v}$ decreases the contribution of the wind component to the light curve (the disk part is independent from $f_{\rm v}$ , see Section~\ref{sec:tde_model}).

The effect of parameter variations on the decline rate of the light curve is shown in Figure \ref{fig:fading_rates_without_diffusion} for $M_6$ and $m_*$ parameters. Here we measure the decline rate as the change in magnitude during 15 rest-frame days after peak. 
We find that the $M_6$ and $m_*$ parameters have the strongest effect on the decline rate, while for all other parameters the variation is less than one percent.
It is also visible that choosing any bandpass in the optical for measuring the light curve has no effect on the decline rate variations.

\subsection{Light curve of a tidal disruption event involving a
white dwarf star}

\begin{figure}
    \centering
    \includegraphics[width=\columnwidth]{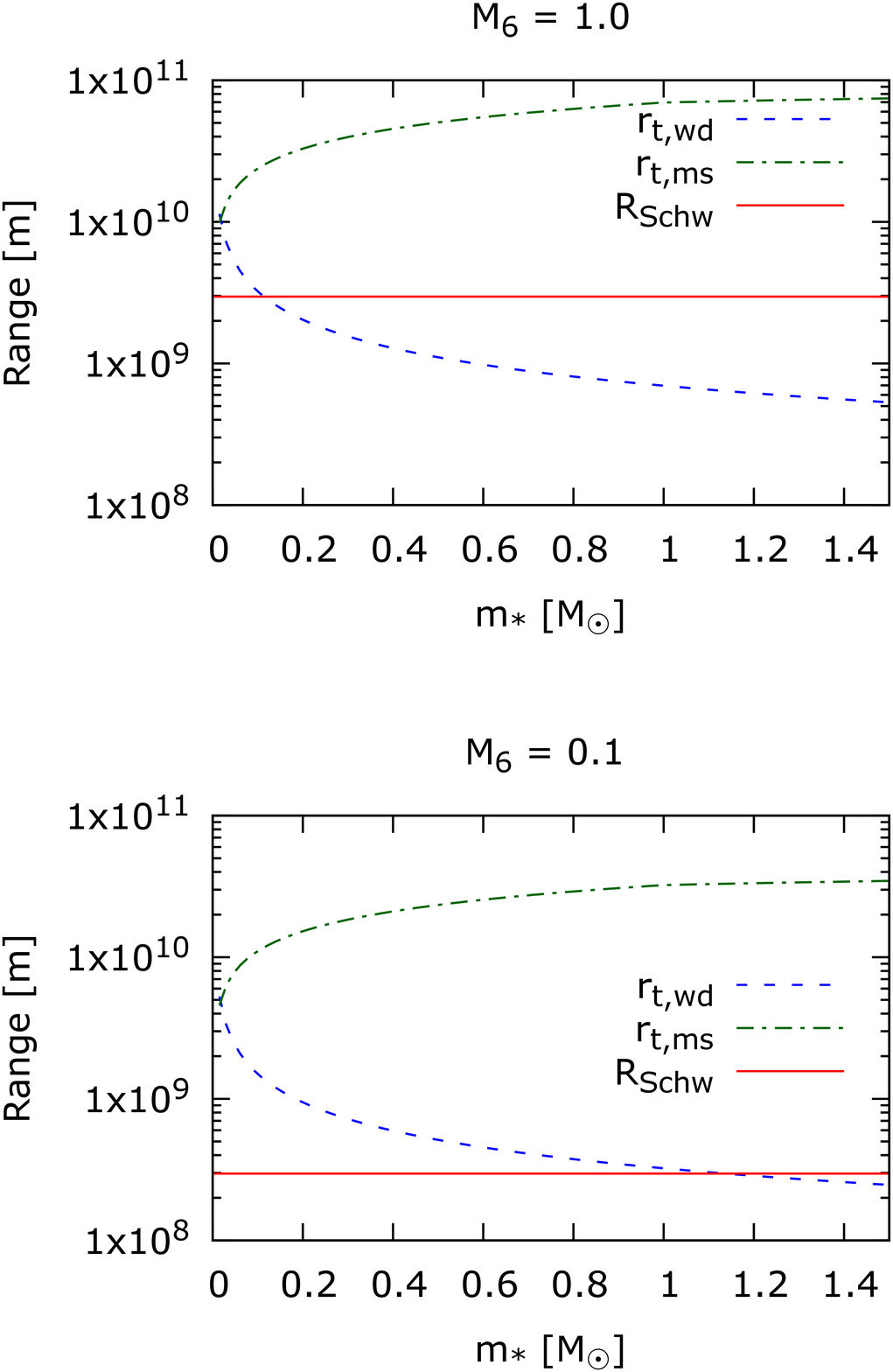}
    \caption{The $r_{\rm t}$ tidal radius of a white dwarf (blue line) and a main sequence star (green line) as a function of the stellar mass. The mass of the black hole is fixed at $10^6 M_\odot$ (upper panel) and at $10^5 M_\odot$ (lower panel). The red line represents the Schwarzschild radius of the BH.}
    \label{fig:wd_possible_mstar}
\end{figure}

When the white dwarf mass-radius relation is applied
to calculate the disruption of a compact star, we have another
important constraint: the $r_{\rm t}$ distance must be higher than the Schwarzschild radius of the BH, $R_{\rm S}$. This constraint results in a strong upper limit on the mass of the white dwarf. While a BH with $M_6 = 1$ allows the disruption of a white dwarf having $m_* \sim 0.1 M_\odot$, this limit increases to $m_* \sim 1.1 M_\odot$ for an $M_6 = 0.1$ BH (see Figure \ref{fig:wd_possible_mstar}). Thus, only relatively massive white dwarfs can be disrupted by $M_6 \leq 0.1$ black holes within the context of the model applied in this paper.  

\begin{figure}
    \centering
    \includegraphics[width=\columnwidth]{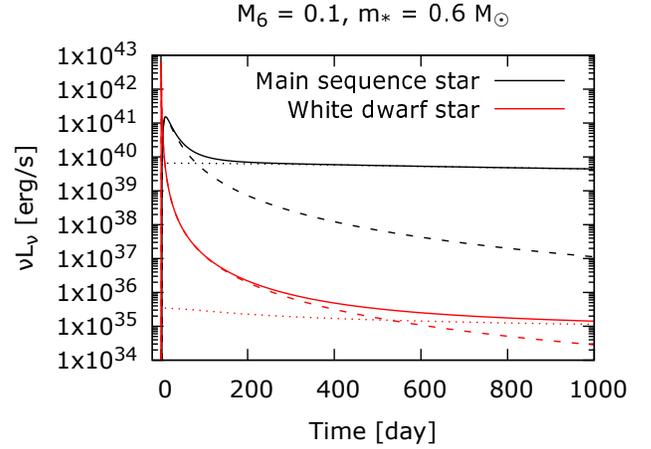}
    \caption{The light curve of a TDE of a main sequence star (black curves) and a white dwarf (red curves). The mass of the black hole is $10^5 M_\odot$ and the star has $0.6 M_\odot$. The radii of the encounters were calculated from their appropriate mass-radius relations, all other parameters are the fiducial ones (Table \ref{Table:initial_parameters}).
    The dashed curves show the luminosity from the wind, while the dotted curves correspond to the disk part. The continuous curves denote the sum of the two components.}
    \label{fig:wd_ms_compare}
\end{figure}

Figure~\ref{fig:wd_ms_compare} intends to compare a TDE of a white dwarf and a main sequence star having the same mass of $m_* = 0.6$. We used L09 accretion rate and corrected $t_{\rm min}$ with $\beta_{\rm d} = 1.8$ for the MS star and $\beta_{\rm d} = 1.2$ for the WD. It is seen that the initial luminosity is higher for the white dwarf and its luminosity declines much faster, while its disk component has much lower luminosity than that from the disruption of a normal main-sequence star.

\section{Testing the model on the AT2019qiz, a TDE candidate}
\label{sec:at2019qiz}
In order to test the model outlined in Sections~\ref{sec:tde_model} and \ref{sec:different_models}, and to demonstrate that our TiDE code produces light curves that are similar to real TDEs, we use the data of a TDE candidate from the Open TDE Catalog\footnote{\tt https://github.com/astrocatalogs}. We applied the following selection criteria to find a suitable object: $i)$ it must have photometric observations available in more than one optical band, the more the better; $ii)$ its photometry must cover both the rising and declining part of the light curve; and $iii)$ it must have published fitting parameters computed with the MOSFiT code. Based on these criteria, our selected candidate is AT2019qiz, a well-observed event. It has extensive observational coverage, and a comprehensive analysis was published recently by \citet{2020MNRAS.499..482N}. 

The downloaded photometric observations were corrected for extinction and redshift, and converted to erg/s/Hz units using the magnitude-flux conversion given by \citet{1998A&A...333..231B}. 

At this step we did not attempt to apply a proper minimization routine and optimize the parameters to get a best-fit result. Instead, our goal was to find a $g$-band light curve having parameters similar to the results from the MOSFiT analysis \citep{2020MNRAS.499..482N} and looks consistent with the data. We used the the time-dependent $f_{\rm out}$ parameter (see Section~\ref{sec:fout}) and included diffusion (Section~\ref{sec:diffusion}). We adopted the L09 accretion rate model (Equation~\ref{eq:mdotfb_l09_all}) and corrected $t_{\rm min}$ (Equation~\ref{eq:tmin_corrected}) with $\beta_{\rm d} = 1.85$. The effect of reprocessing was also taken into account (see Section~\ref{sec:2.2:light_curve}).

\begin{figure*}
    \centering
    \includegraphics[width=2.\columnwidth]{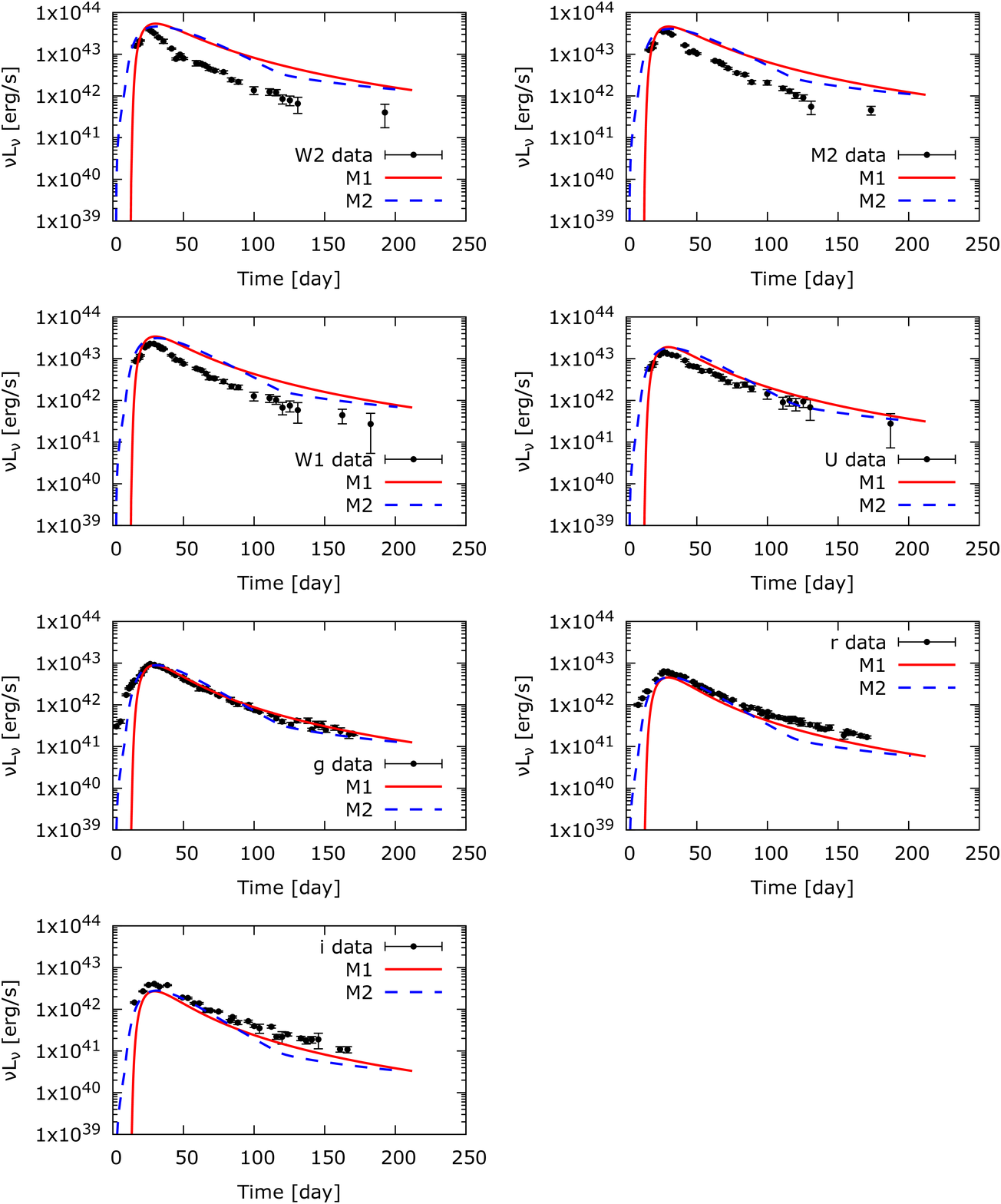}
    \caption{The observed light curves of the AT2019qiz event in different bands (black dots), and the model predictions plotted with red curves. For the model parameters see Table \ref{Table:plotted_parameters}.}
    \label{fig:at2019qiz_best}
\end{figure*}

After finding suitable parameters that produced a light curve similar to the observed $g$-band data, we computed light curves in all other observed bands using the same parameters. The final results are plotted in Figure \ref{fig:at2019qiz_best}, while the parameters are collected in Table \ref{Table:plotted_parameters}, together with the ones found by \citet{2020MNRAS.499..482N} using MOSFiT. 

In Table~\ref{Table:plotted_parameters} we present two slightly different TiDE models. The $g$-band light curve of Model-1 (M1) is most similar to the observed light curve (Figure~\ref{fig:at2019qiz_best}). Since M1 contains a BH having a factor of 5 lower mass than the MOSFiT model, Model-2 (M2) was constructed using the constraint that the BH mass ($M_6$) must agree with that of the MOSFiT model within its uncertainty. Note that in these particular TiDE models reprocessing is almost negligible during the peak and only slightly affects the late part of the light curve in both M1 and M2.

\begin{table}[]
    \centering
    \caption{Model parameters for AT2019qiz}
    
    \begin{tabular}{cccc}
    \hline \hline
    Parameter & M1 & M2 & MOSFiT \\
    \hline
    $M_6$ & 0.15 & 0.7 & $0.8 \pm 0.1$ \\
    $m_*$ & 1.3 & 1.5 & $0.97 \pm 0.04$\\
    $x_*$ & 1.14 & 1.22 & $0.98 \pm 0.03$ \\
    $\eta$ & 0.085 & 0.005 & $0.0058^{+0.0022}_{-0.0006}$\\
    $\beta$ & 0.95 & 0.86 & $0.86 \pm 0.03$\\
    $f_{\rm v}$ & 1.2 & 1.2 & -- \\
    $t_{\rm diff}$ (day) & 4.89 & 4.89 & $5.49^{+0.67}_{-0.7}$ \\
    $\epsilon_{\rm rep}$ & 0.2 & 0.2 & --\\
    \hline

    \end{tabular}
    
    \label{Table:plotted_parameters}
\end{table}

From Figure \ref{fig:at2019qiz_best} we conclude that TiDE produces light curves that are more-or-less similar to MOSFiT (although with a different set of parameters, see below) in the optical bands. In the UV bands the agreement is good at the peak, but after that the TiDE models decline slower than the data. Such a disagreement in the UV is not unexpected, as our simple models use only the blackbody continuum, while the effect of line blanketing is more severe in the UV than in the optical. We discuss the comparison of TiDE and MOSFiT in more details in the next Section.

\section{Discussion}

The TiDE code outlined in Section~\ref{sec:tde_model}, which is based on papers by L09, LR11, SQ09 and GR13, has some differences in its physical assumptions with respect to those of the TDE module in MOSFiT \citep{2019ApJ...872..151M}. This has some important implications regarding the predicted light curve and, as a consequence, on the inferred parameters when the model is compared to observations.

The most important difference is the accretion rate model. MOSFiT uses the hydrodynamically simulated curves from GR13 and scales them with the masses. TiDE adopts the L09 (frozen-in) accretion rate  model, but it was shown that it can produce light curves that are somewhat different from the GR13 results, especially for partial disruptions.

Similar disagreements between some of the estimated parameters can be seen in Table~\ref{Table:plotted_parameters}. The BH mass inferred from our M1 model is almost an order of magnitude lower than the one given by MOSFiT. The other parameters are more-or-less similar to each other, except for the radiative efficiency $\eta$: MOSFiT predicts an unusually low $\eta$ parameter, while our value ($\eta = 0.085$) is in the range which usually adopted in the literature ($0.04 \leq \eta \leq 0.42$) (\citet{1969Natur.223..690L}, \citet{1970Natur.226...64B}) 
Note that since we did not apply a minimization code, we specifically selected an $\eta$ that is similar to the values commonly used in the literature. 

The role of $\eta$ is somewhat controversial in the previously published TDE models. For example, the original model by SQ09 and LR11 is actually independent from this parameter by assuming $r_{\rm L} = 2 r_{\rm p}$ explicitly, which fixes $\eta = 0.01$ (see Section~\ref{sec:tde_model}). We relaxed this constraint by setting $r_{\rm L} = \alpha R_{\rm S}$ with $\alpha$ as a free parameter, resulting in $\eta = 1/(2 \alpha)$. This approach is similar to the one applied in MOSFiT, where $\eta$ is directly tied to the conversion between the emerging luminosity and the mass accretion rate, similar to our Equation~(\ref{eq:Econs}). It is also interesting that the M2 model, which more-or-less fits the data with almost the same BH mass as the MOSFiT model, also needs a very low $\eta$, similar to MOSFiT. However, the stellar mass parameter ($m_*$) in M2 is more than 50 percent larger than the MOSFiT result.

The fact that our M1 model generally predicts lower BH and stellar  masses compared to MOSFiT for the same luminosity level, has probably multiple reasons. One potential cause is the prescription for the accretion rate, which has been shown above to introduce such differences, especially for partial disruptions (which is probably true for this event). Another different assumption is the presence/absence of the super-Eddington wind. In TiDE the super-Eddington wind is responsible for most of the luminosity around the light curve peak. In the MOSFiT implementation, however, the accretion luminosity is limited to being Eddington and below. This could be a fundamental difference in the model assumptions and may lead to significant biases among the best-fit parameters predicted by TiDE and MOSFiT. If the maximum luminosity cannot exceed $L_{\rm Edd}$, then the accretion rate is also limited to be equal or less than $\dot{M}_{\rm Edd}$. Thus, in order to reach the same amount of luminosity as observed, a more massive BH is needed, because $L_{\rm Edd}$ depends linearly on $M$. Since $\eta$ also involved in connecting $L_{\rm Edd}$ and $\dot{M}_{\rm Edd}$ (see Section~\ref{sec:tde_model}), the predicted values of both $M$ and $\eta$ may be strongly affected by these model assumptions. 

Both MOSFiT and TiDE use polytrope models, which are limited compared to real stellar models. The differences between the produced accretion rate in the case of a polytrope and MESA stellar model was studied by \citet{Golightly_1}. Based on their work these differences appear not just in the frozen-in scenario, but in the case of hydrodynamically simulated events for both ZAMS and MAMS MESA models (see their Figure 2 and 4). The incorporation of such more realistic simulations is planned in future versions of TiDE.

\section{Summary}

In this paper we outlined a physically motivated code (TiDE) for predicting the light curve of full tidal disruption events, based on the equations published earlier by SQ09, L09 and LR11.
In Section \ref{sec:different_models}. we introduced new parametrizations to estimate the fallback accretion rate. We compared the resulting accretion rates with the hydrodynamic simulations by GR13, and found that our new models based on the corrected $t_{\rm min}$ definition (Equation~\ref{eq:tmin_corrected}) produces accretion rates for full disruptions that are similar to the simulations presented by GR13 (see Figure \ref{fig:mdotfb_comparsion_4per3} and \ref{fig:mdotfb_comparsion_5per3}).

We applied several modifications and developments on the initial TDE model. It is found that these modifications do not change the light curves significantly, they have relatively minor effects on the general observables (peak luminosity and decline rate).

Adopting a time-dependent (built-in) $f_{\rm out}$ parameter (wind/disk mass ratio) one can reduce the number of free parameters, and it seems to be a plausible assumption: it predicts that the accretion rate decreases with time, thus, the super-Eddington wind sweeps out less and less material. This choice may produce more realistic light curves compared to the assumption of constant $f_{\rm out}$. 

If the super-Eddington wind is optically thick during the early phases, then radiative diffusion may alter the shape of the observed light curve. This effect can be also taken into account via Equation~(\ref{eq:diffusion}). A similar effect for the viscous processes acting during the early phases of a TDE has been implemented in the MOSFiT model as well. Even though $t_{\rm diff}$ is an additional free parameter, which may increase the computational costs of the model, this assumption may result in more realistic light curves (Figure~\ref{fig:diffusion_lcurves}).

We examined the effect of different parameters on the quasi-monochromatic light curves in Section~\ref{sec:light_curves_comparsion} . This can be useful when one attempts to find the best-fit model for a real TDE or TDE candidate. The most important finding is the correlation between the peak of the light curve and the mass of the BH or the disrupted star: the peak brightness in the optical decreases with increasing BH mass, while it increases when the disrupted star is more massive (Figure~\ref{fig:parameter_change}).  

When comparing the decline rates of the inferred light curves as a function of different parameters we found that it mainly depends on the mass of the BH and the star. If the time-dependent $f_{\rm out}$ parameter is applied, then $\eta$ can also produce some variations in the decline rate. The effects of the other parameters on the decline rate are negligible. 

The disruption of different types of objects, namely main-sequence stars and/or white dwarfs, were also studied by using different mass-radius relations. We found that within the context of the model adopted in TiDE, white dwarfs can be disrupted only above a certain mass limit: it is $\sim 0.1$ M$_\odot$ for an $M_6 = 1$ BH, but exceeds 1 M$_\odot$ for $M_6 = 0.1$ BHs. Thus, a lower-mass SMBH can disrupt only relatively massive white dwarfs.

Finally, in Section \ref{sec:at2019qiz}. we showed that TiDE predicts similar light curves to the MOSFiT output, even though the BH mass and the radiative efficiency parameters can be different. More specifically, TiDE predicts less massive BH for the same light curve then MOSFiT, which is mostly attributed to the different prescriptions applied in the two codes for the mass accretion rate: while MOSFiT limits the fallback accretion rates below the Eddington-limit, TiDE uses the super-Eddington wind model to model the peak of the light curve. 

Thus, our intention with this study is to present a model (and a working code) that can be used as an alternative to the MOSFiT module to fit TDE light curves. This may be useful to explore the possible parameter space, and reveal potential biases caused by the model assumptions. TiDE is an object oriented code, and, as a result, it is easy to maintain and extend with different new models such as the accretion model of \citet{nixon21} or the expanding disk model by \citet{2016MNRAS.455..859S}. This way we provide a software environment in which one can compare and test different prescriptions and model assumptions under the same circumstances. The source code of TiDE is publicly available on GitHub\footnote{https://github.com/stermzsofi/TiDE}. Since a TDE is a very complex phenomenon that can occur in widely different configurations and physical conditions, such an approach might get us closer to understanding these exotic and very interesting events. 

\acknowledgments

We express our sincere thanks to the anonymous referee who provided very useful and inspiring comments and suggestions that led to a considerable improvement of this paper. 
This work is supported by the project ``Transient Astrophysical Objects" GINOP 2.3.2-15-2016-00033 of the National Research, Development and Innovation Office (NKFIH), Hungary, funded by the European Union.

\bibliography{tde3_rev}

\begin{thebibliography}{}
\expandafter\ifx\csname natexlab\endcsname\relax\def\natexlab#1{#1}\fi
\providecommand{\url}[1]{\href{#1}{#1}}
\providecommand{\dodoi}[1]{doi:~\href{http://doi.org/#1}{\nolinkurl{#1}}}
\providecommand{\doeprint}[1]{\href{http://ascl.net/#1}{\nolinkurl{http://ascl.net/#1}}}
\providecommand{\doarXiv}[1]{\href{https://arxiv.org/abs/#1}{\nolinkurl{https://arxiv.org/abs/#1}}}

\bibitem[{{Andalman} {et~al.}(2022){Andalman}, {Liska}, {Tchekhovskoy},
  {Coughlin}, \& {Stone}}]{2022MNRAS.510.1627A}
{Andalman}, Z.~L., {Liska}, M. T.~P., {Tchekhovskoy}, A., {Coughlin}, E.~R., \&
  {Stone}, N. 2022, \mnras, 510, 1627, \dodoi{10.1093/mnras/stab3444}

\bibitem[{{Bardeen}(1970)}]{1970Natur.226...64B}
{Bardeen}, J.~M. 1970, \nat, 226, 64, \dodoi{10.1038/226064a0}

\bibitem[{{Bessell} {et~al.}(1998){Bessell}, {Castelli}, \&
  {Plez}}]{1998A&A...333..231B}
{Bessell}, M.~S., {Castelli}, F., \& {Plez}, B. 1998, \aap, 333, 231

\bibitem[{{Bonnerot} \& {Lu}(2020)}]{2020MNRAS.495.1374B}
{Bonnerot}, C., \& {Lu}, W. 2020, \mnras, 495, 1374,
  \dodoi{10.1093/mnras/staa1246}

\bibitem[{{Bonnerot} {et~al.}(2016){Bonnerot}, {Rossi}, {Lodato}, \&
  {Price}}]{2016MNRAS.455.2253B}
{Bonnerot}, C., {Rossi}, E.~M., {Lodato}, G., \& {Price}, D.~J. 2016, \mnras,
  455, 2253, \dodoi{10.1093/mnras/stv2411}

\bibitem[{{Chen} {et~al.}(2009){Chen}, {Madau}, {Sesana}, \&
  {Liu}}]{2009ApJ...697L.149C}
{Chen}, X., {Madau}, P., {Sesana}, A., \& {Liu}, F.~K. 2009, \apjl, 697, L149,
  \dodoi{10.1088/0004-637X/697/2/L149}

\bibitem[{{Chen} {et~al.}(2011){Chen}, {Sesana}, {Madau}, \&
  {Liu}}]{2011ApJ...729...13C}
{Chen}, X., {Sesana}, A., {Madau}, P., \& {Liu}, F.~K. 2011, \apj, 729, 13,
  \dodoi{10.1088/0004-637X/729/1/13}

\bibitem[{{Curd} \& {Narayan}(2019)}]{Curd19}
{Curd}, B., \& {Narayan}, R. 2019, \mnras, 483, 565,
  \dodoi{10.1093/mnras/sty3134}

\bibitem[{{Dai} {et~al.}(2015){Dai}, {McKinney}, \& {Miller}}]{Dai15}
{Dai}, L., {McKinney}, J.~C., \& {Miller}, M.~C. 2015, \apjl, 812, L39,
  \dodoi{10.1088/2041-8205/812/2/L39}

\bibitem[{{Dai} {et~al.}(2018){Dai}, {McKinney}, {Roth}, {Ramirez-Ruiz}, \&
  {Miller}}]{Dai18}
{Dai}, L., {McKinney}, J.~C., {Roth}, N., {Ramirez-Ruiz}, E., \& {Miller},
  M.~C. 2018, \apjl, 859, L20, \dodoi{10.3847/2041-8213/aab429}

\bibitem[{{Dotan} \& {Shaviv}(2011)}]{2011MNRAS.413.1623D}
{Dotan}, C., \& {Shaviv}, N.~J. 2011, \mnras, 413, 1623,
  \dodoi{10.1111/j.1365-2966.2011.18235.x}

\bibitem[{{Evans} \& {Kochanek}(1989)}]{1989ApJ...346L..13E}
{Evans}, C.~R., \& {Kochanek}, C.~S. 1989, \apjl, 346, L13,
  \dodoi{10.1086/185567}

\bibitem[{{Frank} \& {Rees}(1976)}]{1976MNRAS.176..633F}
{Frank}, J., \& {Rees}, M.~J. 1976, \mnras, 176, 633,
  \dodoi{10.1093/mnras/176.3.633}

\bibitem[{{Gallegos-Garcia} {et~al.}(2018){Gallegos-Garcia}, {Law-Smith}, \&
  {Ramirez-Ruiz}}]{Gallegos-Garcia18}
{Gallegos-Garcia}, M., {Law-Smith}, J., \& {Ramirez-Ruiz}, E. 2018, \apj, 857,
  109, \dodoi{10.3847/1538-4357/aab5b8}

\bibitem[{{Gezari}(2021)}]{2021ARA&A..59...21G}
{Gezari}, S. 2021, \araa, 59, \dodoi{10.1146/annurev-astro-111720-030029}

\bibitem[{{Golightly} {et~al.}(2019{\natexlab{a}}){Golightly}, {Coughlin}, \&
  {Nixon}}]{Golightly_2}
{Golightly}, E. C.~A., {Coughlin}, E.~R., \& {Nixon}, C.~J. 2019{\natexlab{a}},
  \apj, 872, 163, \dodoi{10.3847/1538-4357/aafd2f}

\bibitem[{{Golightly} {et~al.}(2019{\natexlab{b}}){Golightly}, {Nixon}, \&
  {Coughlin}}]{Golightly_1}
{Golightly}, E.~C.~A., {Nixon}, C.~J., \& {Coughlin}, E.~R. 2019{\natexlab{b}},
  \apjl, 882, L26, \dodoi{10.3847/2041-8213/ab380d}

\bibitem[{{Guillochon} {et~al.}(2014){Guillochon}, {Manukian}, \&
  {Ramirez-Ruiz}}]{Guillochon14}
{Guillochon}, J., {Manukian}, H., \& {Ramirez-Ruiz}, E. 2014, \apj, 783, 23,
  \dodoi{10.1088/0004-637X/783/1/23}

\bibitem[{{Guillochon} {et~al.}(2018){Guillochon}, {Nicholl}, {Villar},
  {Mockler}, {Narayan}, {Mandel}, {Berger}, \&
  {Williams}}]{2018ApJS..236....6G}
{Guillochon}, J., {Nicholl}, M., {Villar}, V.~A., {et~al.} 2018, \apjs, 236, 6,
  \dodoi{10.3847/1538-4365/aab761}

\bibitem[{{Guillochon} \&
  {Ramirez-Ruiz}(2013{\natexlab{a}})}]{2013ApJ...767...25G}
{Guillochon}, J., \& {Ramirez-Ruiz}, E. 2013{\natexlab{a}}, \apj, 767, 25,
  \dodoi{10.1088/0004-637X/767/1/25}

\bibitem[{{Guillochon} \& {Ramirez-Ruiz}(2013{\natexlab{b}})}]{Guillochon}
---. 2013{\natexlab{b}}, \apj, 767, 25, \dodoi{10.1088/0004-637X/767/1/25}

\bibitem[{{Hayasaki} {et~al.}(2013){Hayasaki}, {Stone}, \& {Loeb}}]{Hayasaki13}
{Hayasaki}, K., {Stone}, N., \& {Loeb}, A. 2013, \mnras, 434, 909,
  \dodoi{10.1093/mnras/stt871}

\bibitem[{{Hayasaki} {et~al.}(2016){Hayasaki}, {Stone}, \& {Loeb}}]{Hayasaki16}
---. 2016, \mnras, 461, 3760, \dodoi{10.1093/mnras/stw1387}

\bibitem[{{Hills}(1975)}]{1975Natur.254..295H}
{Hills}, J.~G. 1975, \nat, 254, 295, \dodoi{10.1038/254295a0}

\bibitem[{{Ivanov} {et~al.}(2005){Ivanov}, {Polnarev}, \&
  {Saha}}]{2005MNRAS.358.1361I}
{Ivanov}, P.~B., {Polnarev}, A.~G., \& {Saha}, P. 2005, \mnras, 358, 1361,
  \dodoi{10.1111/j.1365-2966.2005.08843.x}

\bibitem[{{Kippenhahn} \& {Weigert}(1994)}]{kw94}
{Kippenhahn}, R., \& {Weigert}, A. 1994, {Stellar Structure and Evolution}

\bibitem[{{Komossa}(2015)}]{Komossa15}
{Komossa}, S. 2015, Journal of High Energy Astrophysics, 7, 148,
  \dodoi{10.1016/j.jheap.2015.04.006}

\bibitem[{{Komossa} {et~al.}(2008){Komossa}, {Zhou}, {Wang}, {Ajello}, {Ge},
  {Greiner}, {Lu}, {Salvato}, {Saxton}, {Shan}, {Xu}, \&
  {Yuan}}]{2008ApJ...678L..13K}
{Komossa}, S., {Zhou}, H., {Wang}, T., {et~al.} 2008, \apjl, 678, L13,
  \dodoi{10.1086/588281}

\bibitem[{{Li} {et~al.}(2015){Li}, {Naoz}, {Kocsis}, \&
  {Loeb}}]{2015MNRAS.451.1341L}
{Li}, G., {Naoz}, S., {Kocsis}, B., \& {Loeb}, A. 2015, \mnras, 451, 1341,
  \dodoi{10.1093/mnras/stv1031}

\bibitem[{{Liu} {et~al.}(2022){Liu}, {Mockler}, {Ramirez-Ruiz}, {Yarza},
  {Law-Smith}, {Naoz}, {Melchor}, \& {Rose}}]{2022arXiv220613494L}
{Liu}, C., {Mockler}, B., {Ramirez-Ruiz}, E., {et~al.} 2022, arXiv e-prints,
  arXiv:2206.13494.
\newblock \doarXiv{2206.13494}

\bibitem[{{Lodato} {et~al.}(2009){Lodato}, {King}, \& {Pringle}}]{lodato09}
{Lodato}, G., {King}, A.~R., \& {Pringle}, J.~E. 2009, \mnras, 392, 332,
  \dodoi{10.1111/j.1365-2966.2008.14049.x}

\bibitem[{{Lodato} \& {Rossi}(2011)}]{Lodato11}
{Lodato}, G., \& {Rossi}, E.~M. 2011, \mnras, 410, 359,
  \dodoi{10.1111/j.1365-2966.2010.17448.x}

\bibitem[{{Lu} \& {Bonnerot}(2020)}]{2020MNRAS.492..686L}
{Lu}, W., \& {Bonnerot}, C. 2020, \mnras, 492, 686,
  \dodoi{10.1093/mnras/stz3405}

\bibitem[{{Lynden-Bell}(1969)}]{1969Natur.223..690L}
{Lynden-Bell}, D. 1969, \nat, 223, 690, \dodoi{10.1038/223690a0}

\bibitem[{{MacLeod} {et~al.}(2013){MacLeod}, {Ramirez-Ruiz}, {Grady}, \&
  {Guillochon}}]{2013ApJ...777..133M}
{MacLeod}, M., {Ramirez-Ruiz}, E., {Grady}, S., \& {Guillochon}, J. 2013, \apj,
  777, 133, \dodoi{10.1088/0004-637X/777/2/133}

\bibitem[{{Magorrian} \& {Tremaine}(1999)}]{1999MNRAS.309..447M}
{Magorrian}, J., \& {Tremaine}, S. 1999, \mnras, 309, 447,
  \dodoi{10.1046/j.1365-8711.1999.02853.x}

\bibitem[{{Matsumoto} \& {Piran}(2021)}]{Matsumoto}
{Matsumoto}, T., \& {Piran}, T. 2021, \mnras, 502, 3385,
  \dodoi{10.1093/mnras/stab240}

\bibitem[{{Metzger} \& {Stone}(2016)}]{2016MNRAS.461..948M}
{Metzger}, B.~D., \& {Stone}, N.~C. 2016, \mnras, 461, 948,
  \dodoi{10.1093/mnras/stw1394}

\bibitem[{{Mockler} {et~al.}(2019){Mockler}, {Guillochon}, \&
  {Ramirez-Ruiz}}]{2019ApJ...872..151M}
{Mockler}, B., {Guillochon}, J., \& {Ramirez-Ruiz}, E. 2019, \apj, 872, 151,
  \dodoi{10.3847/1538-4357/ab010f}

\bibitem[{{Nicholl} {et~al.}(2020){Nicholl}, {Wevers}, {Oates}, {Alexander},
  {Leloudas}, {Onori}, {Jerkstrand}, {Gomez}, {Campana}, {Arcavi},
  {Charalampopoulos}, {Gromadzki}, {Ihanec}, {Jonker}, {Lawrence}, {Mandel},
  {Schulze}, {Short}, {Burke}, {McCully}, {Hiramatsu}, {Howell}, {Pellegrino},
  {Abbot}, {Anderson}, {Berger}, {Blanchard}, {Cannizzaro}, {Chen},
  {Dennefeld}, {Galbany}, {Gonz{\'a}lez-Gait{\'a}n}, {Hosseinzadeh}, {Inserra},
  {Irani}, {Kuin}, {M{\"u}ller-Bravo}, {Pineda}, {Ross}, {Roy}, {Smartt},
  {Smith}, {Tucker}, {Wyrzykowski}, \& {Young}}]{2020MNRAS.499..482N}
{Nicholl}, M., {Wevers}, T., {Oates}, S.~R., {et~al.} 2020, \mnras, 499, 482,
  \dodoi{10.1093/mnras/staa2824}

\bibitem[{{Nixon} \& {Coughlin}(2022)}]{2022ApJ...927L..25N}
{Nixon}, C.~J., \& {Coughlin}, E.~R. 2022, \apjl, 927, L25,
  \dodoi{10.3847/2041-8213/ac5118}

\bibitem[{{Nixon} {et~al.}(2021){Nixon}, {Coughlin}, \& {Miles}}]{nixon21}
{Nixon}, C.~J., {Coughlin}, E.~R., \& {Miles}, P.~R. 2021, \apj, 922, 168,
  \dodoi{10.3847/1538-4357/ac1bb8}

\bibitem[{{Phinney}(1989)}]{1989IAUS..136..543P}
{Phinney}, E.~S. 1989, in The Center of the Galaxy, ed. M.~{Morris}, Vol. 136,
  543

\bibitem[{{Piran} {et~al.}(2015){Piran}, {Svirski}, {Krolik}, {Cheng}, \&
  {Shiokawa}}]{2015ApJ...806..164P}
{Piran}, T., {Svirski}, G., {Krolik}, J., {Cheng}, R.~M., \& {Shiokawa}, H.
  2015, \apj, 806, 164, \dodoi{10.1088/0004-637X/806/2/164}

\bibitem[{{Piro} \& {Lu}(2020)}]{Piro}
{Piro}, A.~L., \& {Lu}, W. 2020, \apj, 894, 2, \dodoi{10.3847/1538-4357/ab83f6}

\bibitem[{{Rees}(1988)}]{1988Natur.333..523R}
{Rees}, M.~J. 1988, \nat, 333, 523, \dodoi{10.1038/333523a0}

\bibitem[{{Saxton} {et~al.}(2020){Saxton}, {Komossa}, {Auchettl}, \&
  {Jonker}}]{Saxton20}
{Saxton}, R., {Komossa}, S., {Auchettl}, K., \& {Jonker}, P.~G. 2020, \ssr,
  216, 85, \dodoi{10.1007/s11214-020-00708-4}

\bibitem[{{Shen} {et~al.}(2015){Shen}, {Barniol Duran}, {Nakar}, \&
  {Piran}}]{Shen}
{Shen}, R.~F., {Barniol Duran}, R., {Nakar}, E., \& {Piran}, T. 2015, \mnras,
  447, L60, \dodoi{10.1093/mnrasl/slu183}

\bibitem[{{Shiokawa} {et~al.}(2015){Shiokawa}, {Krolik}, {Cheng}, {Piran}, \&
  {Noble}}]{2015ApJ...804...85S}
{Shiokawa}, H., {Krolik}, J.~H., {Cheng}, R.~M., {Piran}, T., \& {Noble}, S.~C.
  2015, \apj, 804, 85, \dodoi{10.1088/0004-637X/804/2/85}

\bibitem[{{Steinberg} \& {Stone}(2022)}]{2022arXiv220610641S}
{Steinberg}, E., \& {Stone}, N.~C. 2022, arXiv e-prints, arXiv:2206.10641.
\newblock \doarXiv{2206.10641}

\bibitem[{{Stone} \& {Loeb}(2011)}]{2011MNRAS.412...75S}
{Stone}, N., \& {Loeb}, A. 2011, \mnras, 412, 75,
  \dodoi{10.1111/j.1365-2966.2010.17880.x}

\bibitem[{{Stone} {et~al.}(2013){Stone}, {Sari}, \&
  {Loeb}}]{2013MNRAS.435.1809S}
{Stone}, N., {Sari}, R., \& {Loeb}, A. 2013, \mnras, 435, 1809,
  \dodoi{10.1093/mnras/stt1270}

\bibitem[{{Stone} \& {Metzger}(2016)}]{2016MNRAS.455..859S}
{Stone}, N.~C., \& {Metzger}, B.~D. 2016, \mnras, 455, 859,
  \dodoi{10.1093/mnras/stv2281}

\bibitem[{{Strubbe} \& {Quataert}(2009)}]{Strubbe09}
{Strubbe}, L.~E., \& {Quataert}, E. 2009, \mnras, 400, 2070,
  \dodoi{10.1111/j.1365-2966.2009.15599.x}

\bibitem[{{van Velzen} {et~al.}(2020){van Velzen}, {Holoien}, {Onori}, {Hung},
  \& {Arcavi}}]{2020SSRv..216..124V}
{van Velzen}, S., {Holoien}, T. W.~S., {Onori}, F., {Hung}, T., \& {Arcavi}, I.
  2020, \ssr, 216, 124, \dodoi{10.1007/s11214-020-00753-z}

\bibitem[{{van Velzen} {et~al.}(2019{\natexlab{a}}){van Velzen}, {Stone},
  {Metzger}, {Gezari}, {Brown}, \& {Fruchter}}]{vanVelzen19}
{van Velzen}, S., {Stone}, N.~C., {Metzger}, B.~D., {et~al.}
  2019{\natexlab{a}}, \apj, 878, 82, \dodoi{10.3847/1538-4357/ab1844}

\bibitem[{{van Velzen} {et~al.}(2011){van Velzen}, {Farrar}, {Gezari},
  {Morrell}, {Zaritsky}, {{\"O}stman}, {Smith}, {Gelfand}, \&
  {Drake}}]{2011ApJ...741...73V}
{van Velzen}, S., {Farrar}, G.~R., {Gezari}, S., {et~al.} 2011, \apj, 741, 73,
  \dodoi{10.1088/0004-637X/741/2/73}

\bibitem[{{van Velzen} {et~al.}(2019{\natexlab{b}}){van Velzen}, {Gezari},
  {Cenko}, {Kara}, {Miller-Jones}, {Hung}, {Bright}, {Roth}, {Blagorodnova},
  {Huppenkothen}, {Yan}, {Ofek}, {Sollerman}, {Frederick}, {Ward}, {Graham},
  {Fender}, {Kasliwal}, {Canella}, {Stein}, {Giomi}, {Brinnel}, {van Santen},
  {Nordin}, {Bellm}, {Dekany}, {Fremling}, {Golkhou}, {Kupfer}, {Kulkarni},
  {Laher}, {Mahabal}, {Masci}, {Miller}, {Neill}, {Riddle}, {Rigault},
  {Rusholme}, {Soumagnac}, \& {Tachibana}}]{2019ApJ...872..198V}
{van Velzen}, S., {Gezari}, S., {Cenko}, S.~B., {et~al.} 2019{\natexlab{b}},
  \apj, 872, 198, \dodoi{10.3847/1538-4357/aafe0c}

\bibitem[{{van Velzen} {et~al.}(2021){van Velzen}, {Gezari}, {Hammerstein},
  {Roth}, {Frederick}, {Ward}, {Hung}, {Cenko}, {Stein}, {Perley}, {Taggart},
  {Foley}, {Sollerman}, {Blagorodnova}, {Andreoni}, {Bellm}, {Brinnel}, {De},
  {Dekany}, {Feeney}, {Fremling}, {Giomi}, {Golkhou}, {Graham}, {Ho},
  {Kasliwal}, {Kilpatrick}, {Kulkarni}, {Kupfer}, {Laher}, {Mahabal}, {Masci},
  {Miller}, {Nordin}, {Riddle}, {Rusholme}, {van Santen}, {Sharma}, {Shupe}, \&
  {Soumagnac}}]{2021ApJ...908....4V}
{van Velzen}, S., {Gezari}, S., {Hammerstein}, E., {et~al.} 2021, \apj, 908, 4,
  \dodoi{10.3847/1538-4357/abc258}

\bibitem[{{Wang} {et~al.}(2012){Wang}, {Zhou}, {Komossa}, {Wang}, {Yuan}, \&
  {Yang}}]{2012ApJ...749..115W}
{Wang}, T.-G., {Zhou}, H.-Y., {Komossa}, S., {et~al.} 2012, \apj, 749, 115,
  \dodoi{10.1088/0004-637X/749/2/115}

\bibitem[{{Wang} {et~al.}(2011){Wang}, {Zhou}, {Wang}, {Lu}, \&
  {Xu}}]{2011ApJ...740...85W}
{Wang}, T.-G., {Zhou}, H.-Y., {Wang}, L.-F., {Lu}, H.-L., \& {Xu}, D. 2011,
  \apj, 740, 85, \dodoi{10.1088/0004-637X/740/2/85}

\bibitem[{{Xiang-Gruess} {et~al.}(2016){Xiang-Gruess}, {Ivanov}, \&
  {Papaloizou}}]{2016MNRAS.463.2242X}
{Xiang-Gruess}, M., {Ivanov}, P.~B., \& {Papaloizou}, J.~C.~B. 2016, \mnras,
  463, 2242, \dodoi{10.1093/mnras/stw2130}

\bibitem[{{Yang} {et~al.}(2013){Yang}, {Wang}, {Ferland}, {Yuan}, {Zhou}, \&
  {Jiang}}]{2013ApJ...774...46Y}
{Yang}, C.-W., {Wang}, T.-G., {Ferland}, G., {et~al.} 2013, \apj, 774, 46,
  \dodoi{10.1088/0004-637X/774/1/46}

\bibitem[{{Zhuang} \& {Shen}(2021)}]{2021JHEAp..32...11Z}
{Zhuang}, J., \& {Shen}, R.-F. 2021, Journal of High Energy Astrophysics, 32,
  11, \dodoi{10.1016/j.jheap.2021.06.001}

\end{thebibliography}

\end{document}